%% file: main.tex
\documentclass[journal,twcolumn,final]{IEEEtran}

\usepackage{textpos}
\newcommand\copyrighttext{
	\smaller This article has been accepted for publication in IEEE Transactions on Network and Service Management.  
	
}
\newcommand\copyrightnotice{%
	\begin{tikzpicture}[remember picture,overlay]
		\node[anchor=north,yshift=-15pt] at (current page.north)  {\copyrighttext}
		;
	\end{tikzpicture}%
}

\usepackage{setspace}

\usepackage{bbm}
\usepackage{amsmath}
\usepackage{bm}
\usepackage{mdframed}
\usepackage{amsthm}

\usepackage[utf8]{inputenc}
\usepackage{graphicx}
\usepackage[colorinlistoftodos]{todonotes}
\usepackage{amssymb}
\usepackage{longtable}
\usepackage[colorlinks=true, allcolors=blue]{hyperref}

\usepackage[noadjust]{cite}
\allowdisplaybreaks
\usepackage{lineno}
\usepackage{lipsum}
\usepackage{multicol}
\usepackage{subcaption}

\usepackage{multirow}
\usepackage{array}

\usepackage{xcolor}

\usepackage[vlined,linesnumbered,ruled,resetcount]{algorithm2e}
\SetArgSty{textnormal}

\usepackage{bbm}

\SetKwInput{KwInput}{Input}                
\SetKwInput{KwOutput}{Output}              
\SetKwInput{KwInitialize}{Initialize}                
\SetKwInput{myproc}{Procedure}{}{}
\usepackage{booktabs}
\usepackage{caption}
\usepackage{float}
\usepackage{capt-of}
\usepackage{arydshln}
\usepackage{soul,xcolor}
\usepackage{tabularx}

\usepackage{arydshln}
\usepackage[flushleft]{threeparttable}

\newcommand{\rev}[1]{{\color{black}#1}}

\title{Deep Reinforcement Learning for Orchestrating Cost-Aware Reconfigurations of vRANs}

\author{\IEEEauthorblockN{Fahri Wisnu Murti, Samad Ali, George Iosifidis, Matti Latva-aho}\\
\thanks{
Fahri Wisnu Murti, Samad Ali and Matti Latva-aho are with Centre for Wireless Communications, University of Oulu, Finland. 
George Iosifidis is with Delft University of Technology, Netherlands. 

	{This research has been supported by the Academy of Finland, 6G Flagship program under Grant 346208. F. W. Murti also would like to acknowledge the support of Nokia Foundation. }
}%
}

\IEEEoverridecommandlockouts
\begin{document}
\sloppy

\maketitle
\thispagestyle{plain}
\pagestyle{plain}
\begin{abstract}
\rev{Virtualized Radio Access Networks (vRANs) are fully configurable and can be implemented at a low cost over commodity platforms to enable network management flexibility. In this paper, a novel vRAN reconfiguration problem is formulated to jointly reconfigure the functional splits of the base stations (BSs), locations of the virtualized central units (vCUs) and distributed units (vDUs), their resources, and the routing for each BS data flow. The objective is to minimize the long-term total network operation cost while adapting to the varying traffic demands and resource availability. In the ﬁrst step, testbed measurements are performed to study the relationship between the traffic demands and computing resources, which reveals high variance and depends on the platform and its load. Consequently, finding the perfect model of the underlying system is non-trivial. Therefore, to solve the proposed problem, a deep reinforcement learning (RL)-based framework is proposed and developed using model-free RL approaches. Moreover, the problem consists of multiple BSs sharing the same resources, which results in a multi-dimensional discrete action space and leads to a combinatorial number of possible actions. To overcome this curse of dimensionality, action branching architecture,  which is an action decomposition method with a shared decision module followed by neural network is combined with Dueling Double Deep Q-network (D3QN) algorithm. Simulations are carried out using an O-RAN compliant model and real traces of the testbed. Our numerical results show that the proposed framework successfully learns the optimal policy that adaptively selects the vRAN configurations, where its learning convergence can be further expedited through transfer learning even in different vRAN systems. It also offers significant cost savings by up to 59\% of a static benchmark, 35\% of Deep Deterministic Policy Gradient with discretization, and 76\% of non-branching D3QN.}
\end{abstract}


\IEEEpeerreviewmaketitle

\copyrightnotice

\vspace{-2mm}
\input{Introduction.tex}

\vspace{-1mm}
\input{Related_Work.tex}

\input{Model.tex}

\vspace{-2mm}
\input{Algorithms.tex}


\vspace{-2mm}
\input{Simulations.tex}

\vspace{-2mm}
\input{Conclusions.tex}



{\scriptsize
	\vspace{-1mm}
	\bibliographystyle{IEEEtran}
	\bibliography{IEEEabrv,ref_journal}
}

\end{document}

%% file: Introduction.tex
\section{Introduction}
\subsection{Motivation}
Virtualization has become one of the most promising technologies for accommodating the increased service demands with diverse requirements at a reasonable cost in cellular networks \cite{bonati_survey}. The latest effort of this idea is virtualizing the radio access networks (vRANs) by replacing the hardware-based legacy RANs with softwarized RANs \cite{nokia5g, nokia_anyhaul, openvran_nec}. Incorporated with Open RAN, vRANs can be fully conﬁgurable and deployed across heterogeneous platforms such as commodity servers and small embedded devices. Another exciting feature of vRANs is that it enables the baseband functions (BBU) of each \rev{base station} (BS) to be disaggregated, hosted at the virtualized distributed units (vDUs) and central units (vCUs), and executed as virtual machine (VM) instances or light-weight containers over geo-distributed locations.
This paradigm shift brings unprecedented flexibility to RAN operations, mitigates vendor lock-in, offers fast deployment and potentially reduces operational expenses \cite{openvran_nec}. Therefore, it is not surprising that many standardization bodies envision the virtualization for their future RANs, such as O-RAN \cite{oran_architecture} and 5G+ RAN \cite{3gpp_rel16}.

Nevertheless, the expansive deployment of vRANs is still hindered by complex configuration options, which introduce new network management challenges in deploying cost-efficient vRAN configurations while serving the traffic demands.   
In particular, the operators need to decide the \textit{functional splits} of the BSs to determine which BS functions are deployed at the vDUs and which are at the vCUs. Each choice of these splits has a different delay requirement, consumes different computing resources for the vDUs/vCUs, and generates a different data load over the xHaul links\footnote{\rev{The paths connecting a core network (EPC) to vCUs, vCUs to vDUs, and vDUs to radio units (RUs) are defined as backhaul (BH), midhaul (MH), fronthaul (FH), respectively, and the integration of these elements is called Crosshaul/xHaul transport network.} }.
Moreover, the vDUs/vCUs are executed on top of commodity platforms as VM instances or containers; hence, the operators need to allocate the \textit{virtualized resources} (e.g., CPUs, memory) for them. 
There are also several candidate deployment locations for each vDU/vCU, possibly with different hosting machines, and this creates the \textit{placement problem} in determining their optimal locations and platforms. 
At the same time, each placement location is associated with different eligible \textit{routing paths} to transfer the data flow of the BSs, which incur particular delays and costs.
Consequently, these issues create a challenging coupling among the BS splits, placement and allocated resources for the vDUs/vCUs, and routing for each BS data flow.

Meanwhile, the suitability of the vRAN configurations is highly affected by the network properties such as traffic demands and resource availability (e.g., computing and xHaul link capacity) \cite{vran_optimal_murti2}, which might change over time, often in an unpredictable fashion\footnote{This is particularly common for resource availability/costs in shared infrastructures or traffic and channel conditions in small cell networks \cite{pachos_caching}.}. Thus, deploying static conﬁgurations for a long time might result in resource overprovisioning or even declined traffic demands. Resource overprovisioning occurs when the allocated resources are higher than the actual resource utilization. The declined demands can be triggered by insufﬁcient allocated resources (underprovisioning) and constraint violation. And these can render substantial performance degradation and high operating expenditures. Therefore, it is essential to dynamically select the vRAN conﬁgurations to adapt to varying traffic demands and resource availability.

On the other hand, orchestrating the dynamic configurations of vRANs is a non-trivial endeavor. The reconfiguration decisions must be enforced before the actual traffic demands of the BSs are observed. Albeit reconfiguring the vRAN system at runtime is practically possible \cite{adaptive_vran_alba}, it might also induce additional costs and disrupt network operations during the live migration of the VM instances. Consequently, any reconfiguration activity needs to be performed prudently to ensure that it is beneficial both in terms of cost and performance.
However, designing such an intelligent approach also has technical issues since the software-based vRAN system substantially differs from hardware-based legacy RANs. The takeaway from \rev{our testbed measurements (details in Sec. \ref{sec:simulation})} and prior experimental studies (cf., \cite{vranai_journal,bayes_vran_journal}) is that, unlike legacy RANs, the underlying system of vRANs is complex, poly-parametric and has platform-dependent performance. Hence, adopting traditional control policies, which needs perfect knowledge of the underlying system to model and solve the problems, is unrealistic in practice.

Motivated by \rev{the challenges above and our measurement insights (details in Sec. \ref{sec:simulation})}, we propose and study \textit{a fresh vRAN reconfiguration problem}, where it jointly reconfigures the splits of the BSs, resources and locations of the vDUs and vCUs, and the routing of each BS data flow to minimize the long-term total network operation \rev{cost}. The key idea is to model this problem as reinforcement learning (RL) and \rev{develop \textit{a learning-based framework}, namely \textbf{L}earning-based \textbf{A}utomated \textbf{R}econfiguration for \textbf{v}RANs (LARV), to solve the problem with minimal assumptions about the system.}

\vspace{-2mm}
\subsection{Contributions and Methodology}

We firstly build a prototype implementing the centralized-RAN (C-RAN) system using software-based srsRAN \cite{srslte} in two different platforms to collect measurements regarding the relations between traffic demands and resource utilization. The findings reveal that the relations vary with the demands and, importantly, have high variance and dependence on the platform, platform load\footnote{\rev{The relations heavily rely on the types of platforms that host the BBU. It also depends on platform load (e.g., when vRAN workload shares the same platform with other applications or workloads such as edge computing, data analytics, etc.); see Sec. \ref{sec:simulation} and \cite{vranai_journal, edgebol, concordia} for details.} }, and many latent factors. These inhibit adopting general assumptions of the underlying system (e.g., linear) and traditional mathematical tool-based policies. Then, we propose a new cost model accounting for resource overprovisioning, instantiation and reconfiguration, and the declined traffic demands, representing the virtualized resource management in vRANs. This model also considers different computing and routing costs for each split and platform location.
Further, we model our vRAN system following the latest proposal of O-RAN architecture \cite{oran_architecture}. We consider a vRAN system with multiple BSs and define its operation as a time-slotted system, where each slot has arbitrary incoming traffic demands and resource availability. At each time slot, LARV takes an action that selects the vRAN configurations, then reconfigures the system when the selected are different from the last configurations or preserves them if the selected configurations are the same. LARV expects to receive a reward signal from the system that assesses the quality of each selected action. This sequential decision-making is formulated as Markov decision process (MDP), which is also an RL problem.

In our solution, LARV is developed using model-free RL with deep neural network architecture. LARV considers the vRAN system as a black-box environment and does not make any particular assumptions about the underlying system state and state transition probability distribution. Since the formulated RL problem has a semi-continuous state space and discrete action space, we propose a Dueling Double Deep Q-network (D3QN)-based approach \cite{dueling_dqn}, in which the learning step is based on Double Q-learning \cite{ddqn}. However, the system has multiple BSs that share the same resources with highly coupled configuration decisions. As a result, the RL formulation renders a multi-dimensional action space, which exhibits combinatorial growth of the number of possible actions. In order to overcome the curse dimensionality, the proposed D3QN is incorporated with action branching \cite{bdq}, an action decomposition method that decomposes the multi-dimension action into sub-actions and utilizes shared decision module followed by neural network branches. However, the initial action branching proposed in \cite{bdq} focused on sub-actions with the same dimensional size, which can not be directly applied to our problem. Here, we adapt it; hence each sub-action dimension can vary but still exhibits a linear growth of the total neural network outputs (estimated actions) with the increase of action dimensionality while maintaining the shared decision.

We conduct a battery of tests using an O-RAN compliant model and real traces collected from the testbed. We evaluate the training behavior and long-term total network cost during online operation under various scenarios. Our numerical results reveal that LARV successfully learns the optimal policy to select an action that controls the vRAN configurations, \rev{where its learning convergence can be accelerated via transfer learning even in different vRAN systems}. \rev{Moreover}, LARV offers considerable cost savings by up to 59\% of a static benchmark, 35\% of Deep Deterministic Policy Gradient (DDPG) with discretization, and 76\% of distributed non-branching D3QN.
Our contributions can be summarized:
\begin{itemize}
	\item We propose and study a new vRAN reconfiguration problem, where it jointly reconfigures splits of the BSs, resources and locations of the vDUs/vCUs, routing for each BS flow.
	\item We carefully model our vRAN system based on the latest proposals of O-RAN architecture and propose a comprehensive cost model. The model takes resource overprovisioning, instantiation and reconfiguration and the declined demands costs into account. It also captures platform/split-dependent computing and routing costs.
	\item We develop a learning-based framework to solve the proposed vRAN reconfiguration problem. It is tailored from D3QN and an action branching architecture to tackle the multi-dimensional and large action space inherited from our RL problem with linear growth of the neural network outputs.
	\item We conduct extensive trace-driven simulations and analyze the performance of LARV under various scenarios during the training process and online operation. 
	
\end{itemize}

The rest of this paper is organized as follows. \rev{Sec. \ref{sec:relatedwork} discusses our contributions with respect to prior works.} In Sec. \ref{sec:model}, the architecture background and model used for our vRAN system are presented. The reconfiguration problem is also formulated in this section, including the raised trade-offs. In Sec. \ref{sec:algo}, we discuss how to design the proposed learning algorithm. The detailed experiment setups, \rev{testbed measurement insights}, and simulation results are presented in Sec. \ref{sec:simulation}. Finally, our paper is concluded in Sec. \ref{sec:conclusion}.

%% file: Related_Work.tex
\rev{\section{Related Work} \label{sec:relatedwork}}

Recent works have studied various vRAN orchestration problems, and we can classify them into \textit{i)} those that rely on models to optimize the configurations and \textit{ii)} model-free approaches that utilize offline training data and \textit{iii)} RL methods. The examples of the first point include \cite{andres_fluidran_joint,ojaghi_oran_mec} that optimize the vRAN functional splits with \rev{multi-access edge computing (MEC)} services, \cite{vran_optimal_murti2} that considers the functional split problem with multiple servers for hosting the vCUs, and \cite{placeran} that further expands it to several candidate servers to place the vCUs/vDUs. Albeit they have optimized various configurations in vRANs, they aimed for offline network designs, and the implication of varying conditions from traffic demands and resource availability is still not examined. The studies of model-based approaches that consider varying conditions include altering the functional splits at runtime to maximize the users' throughput \cite{dynamic_split_alba} and revenue  \cite{alba_cost_split} and to minimize inter-cell interference and FH utilization \cite{flex5g}. Another example in \cite{cares} aimed to control radio/computing scheduling to maximize the served traffic subject to a BS computing capacity. However, they \cite{dynamic_split_alba, alba_cost_split, flex5g,cares} still did not study where to place and how much the allocated resources are for the vDUs/vCUs, although these configurations play crucial role in a vRAN system. Moreover, such model-based approaches can be impractical as they heavily rely on fine-tuning models for specific scenarios and underlying system assumptions. And a vRAN system is network and platform-dependent, where the models can be unknown in practice.
On the other hand, model-free approaches employing machine learning (ML) have been increasingly popular in tackling complex problems in mobile networks. Particularly, approaches that employ function approximation of performance metrics, e.g., via neural networks, can offer satisfactory performance amidst many unknown system parameters \cite{ali20206g}. For instance, the authors in \cite{matoussi_userslice} have developed a deep supervised learning framework for allocating radio resources and functional split for each user. Such supervised learning can deliver well-achieved performance as long as there are high-quality labeled datasets, e.g., optimal labels. However, the optimal labels are often not be available in vRAN problems. Hence, those that do not require labeled datasets, such as contextual bandit and \rev{full} RL formulations, can be leveraged. The authors in \cite{vranai_journal} have tailored a deep learning-based framework to solve the contextual bandit problem of managing the interplay between computing and radio resources. The other contextual bandits in \cite{bayes_vran_journal} and \cite{michail_bayes_oran} utilize a data-efficient algorithm, Bayesian online learning for an energy-aware BS in a vRAN system. These approaches offer remarkable performance with the condition that the current context observation must not be affected by the previous actions, i.e., it only includes exogenous parameters.

Otherwise, a \rev{full} RL formulation is required when the current observation, e.g., state, is influenced by the previous actions. Recent work in \cite{rl_oran} has brought the importance of a model-free RL formulation by utilizing Q-learning and SARSA algorithms to optimize the functional split selections for an energy-efficient O-RAN. However, when the state-action space of the RL problem is large, such approaches become inefficient. Therefore, a deep RL paradigm can be utilized to tackle such issue by using neural network architecture to approximate the state-action function. \rev{Some interesting examples} are \cite{bonati_oran} and \cite{bonati_coloran} that have developed xApps for controlling RAN slicing, scheduling and online model training using the Proximal Policy Optimization algorithm. 
In \cite{cdrs}, the authors also have solved the functional split problem by proposing a chain rule-based stochastic policy and approximate it with sequence-to-sequence model.
Our recent work in \cite{lofv} has proposed an RL-based framework using a combination of Deep Q-Network (DQN) and a regressor to dynamically reconfigure the functional split and its required computing resources. However, it was still limited to a single BS and did not include computing and link resource sharing among the BSs.

\rev{Although the mentioned works have solved various adaptive vRAN orchestration problems, they mainly focused on controlling functional splits (e.g., \cite{dynamic_split_alba, alba_cost_split, flex5g,cares, rl_oran}), RAN slicing (e.g., \cite{bonati_coloran,bonati_oran}) and radio/computing scheduling (e.g., \cite{matoussi_userslice,vranai_journal,bayes_vran_journal, michail_bayes_oran,  rl_oran, bonati_coloran, bonati_oran}). On the other hand, the joint reconfiguration between functional splits of the BSs, the virtualized resource allocation and placement for the vCUs/vDUs over geo-distributed cloud platforms, and the routing, along with the impacts of altering such configurations at runtime, are \textit{hitherto unexplored}. Here, we aim to fill a gap by tackling this reconfiguration problem using model-free RL that makes minimal assumptions about the system. Since the problem also consists of multiple BSs with highly coupled configurations, the RL formulation renders a dimensional explosion in the state space and action space, making the available vRAN orchestration frameworks unsuitable. To solve this challenging dimensionality issue, we develop LARV, a novel vRAN orchestration framework based on deep RL, from the incorporation of action branching with D3QN.}

%% file: Model.tex
\section{System Model and Problem Formulation} \label{sec:model}
\subsection{Background and Model}
\begin{figure}[t!]
	\centering
	\includegraphics[width=0.48\textwidth]{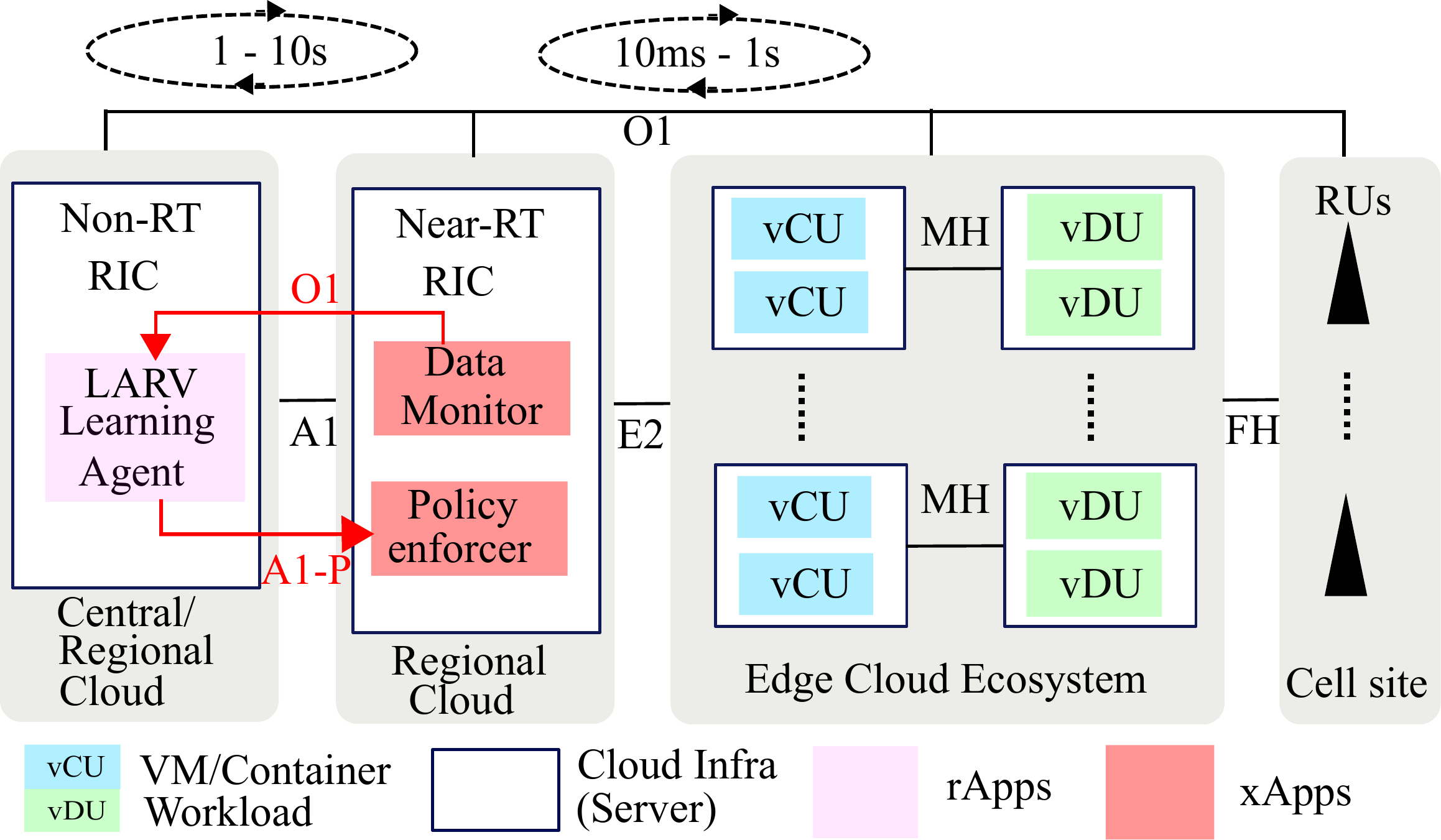}   
	\caption{\rev{\small{O-RAN compliant system architecture adopted in our model.}}}
	\label{fig:oran}
\end{figure}

\rev{We model our vRAN system following the latest proposals of O-RAN architecture \cite{oran_architecture}, where the high-level architecture is illustrated in Fig. \ref{fig:oran}. The model adopts O-RAN key principles that include disaggregation, virtualization, open interfaces, and intelligent control \cite{polese_understanding_oran}. The protocol stacks (or functions) of each BS can be disaggregated through the functional split and, further, virtualized as the vCU and vDU (connected to an RU). Hence, a BS corresponds to 4G eNodeB or 5G gNodeB comprising a vCU, vDU, and RU. The vCU and vDU can be executed as VM instances or containers across geo-distributed edge cloud infrastructures, which may share with other workloads. Then, the intelligent control is realized through RAN Intelligent Controllers (RICs), which can run routine optimization and orchestration through closed-loop control. O-RAN has specified two RICs: i) Non-Real-Time (Non-RT) RIC and ii) Near-Real-Time (Near-RT) RIC. The Non-RT RIC, which integrates with the network orchestrator,  operates on a time scale longer than 1 s, while the Near-RT RIC operates with a time scale between 10 ms and 1s. The Non-RT RIC supports applications, called rApps, that support RAN optimization and operations such as policy guidance, configuration management, etc. While the Near-RT RIC includes applications called xApps that can be used to perform radio resource management.
Then, LARV is to be implemented in the learning agent as an rApp in the Non-RT RIC in the system orchestrator of O-RAN and enforces a policy at every period of $n \!=\! 1, ..., N$ to control the reconfigurations of BSs. 
The optimal policy at every time $n$ depends on the input observation (state), which is provided at the beginning of each period by the BSs via the O1 interface. }


\begin{figure}[t!]
	\centering
	\includegraphics[width=0.48\textwidth]{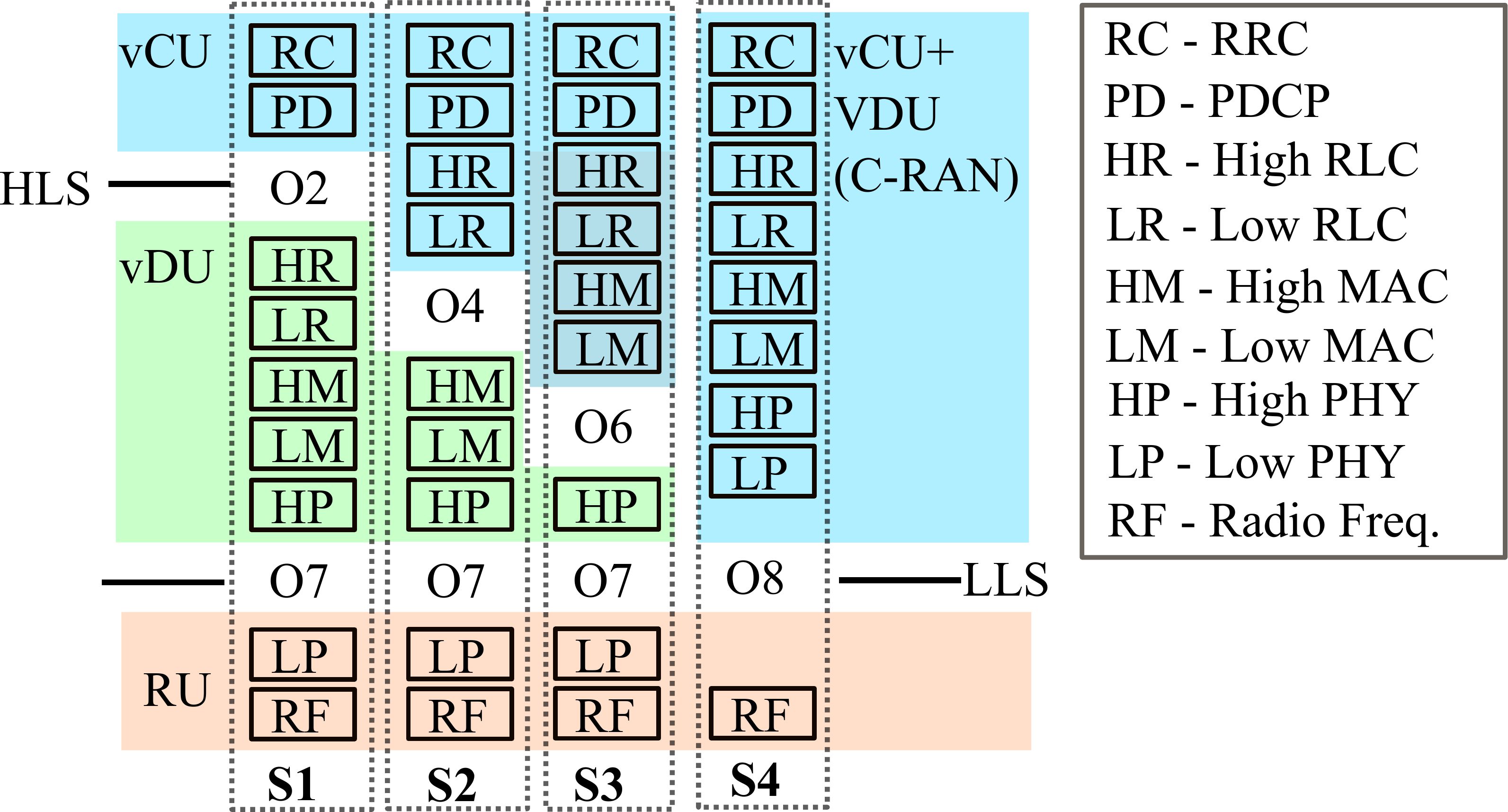}   
	\caption{\small The functional splits applied in our vRAN model. S1, S2 and S3 are envisioned in O-RAN architecture proposals, while S4 are legacy C-RAN. }
	\label{fig:split_option}
\end{figure}


\rev{Next, we illustrate the functional split options used in our model in Fig. \ref{fig:split_option} and present their requirements in Table \ref{table:3gpp_split}. As suggested by O-RAN \cite{oran_architecture}, we consider Option 7.x (O7) and Option 8 (O8) for the Low Layer Split (LLS) between the vDU and RU. The High Layer Split (HLS) between the vCU and vDU can use Option 2 (O2), which is currently the most feasible split to be implemented. We also consider Option 4 (O4) and Option 6 (O6), which have been well standardized \cite{oran_architecture,3gpp_rel16} and experimentally validated \cite{dynamic_split_alba}, to encourage further RAN flexibility.
Therefore, following HLS and LLS, we denote four choices of functional splits: \textit{Split 1 (S1)} implements O2 for the HLS and O7 for the LLS; \textit{Split 2 (S2)} uses O4 for the HLS and O7 for the LLS; \textit{Split 3 (S3)} adopts O6 for the HLS and O7 for the LLS; and \textit{Split 4 (S4)} is the legacy C-RAN system, which implements Option 8 (O8), i.e., all the BS functions are executed as an integrated vDU/vCU except RF functions (at the RU). We define a set of these four possible splits as $\mathcal{I} = \{ \text{S1,S2,S3,S4}\}$. }

\begin{table}[t] \centering
	\begin{threeparttable}
		\begin{small}
			\begin{tabular}{@{}lllll@{}}\toprule
				\textbf{}& \textbf{Split Point} & \textbf{Load} &\textbf{Max} & \textbf{Delay Req.}  
				\\ \midrule
				{O1 } &  RRC - PDCP &   $\lambda$ & $ 4$   & $10$ ms         
				\\ \hdashline
				{O2$^*$ } &  PDCP - High RLC &  $\lambda$ & $ 4$  & $10$ ms
				\\ \hdashline
				{O3 } &  High RLC - Low RLC & $\lambda$ & $ 4$    & $10$ ms
				\\ \hdashline
				{O4$^*$ } &       Low RLC - High MAC & $\lambda$ & $ 4$   & $1$ ms         
				\\ \hdashline
				{O5 } & High MAC - Low MAC  & $\lambda$ &  {$ 4$}  & $1$ ms
				\\ \hdashline
				{O6$^*$ } &  Low MAC - High PHY  & $1.02\lambda$+0.5 & {$ 4.13 $}     & $0.25$ ms
				\\ \hdashline
				{O7$^\dagger$ } & High PHY - Low PHY &   {$ 10.1$} &  {$ 10.1$}   & $0.25$ ms
				\\ \hdashline
				{O8$^\dagger$ } &   Low PHY - RF  & {$157.3$}  &{$157.3$}   & $0.25$ ms \\ 
				\bottomrule
			\end{tabular}
			\begin{tablenotes}
				\item Note: $^*$ is applied options for HLS and $^\dagger$ is applied options for LLS. The data load is in Gbps.
			\end{tablenotes}
		\end{small}
	\end{threeparttable}
	\caption{\small The functional split options and their requirements based on 3GPP nomenclature when the traffic demand is $\lambda$ Gbps. The requirements are tailored by following settings: 100 MHz bandwidth, 256 QAM, 32 antenna ports and 8 MIMO layers. The achievable data rate is up to 4 Gbps.}
	\label{table:3gpp_split}
\end{table}

%
\begin{table*}[t] \centering
	\begin{threeparttable}
		\begin{small}
			\begin{tabular}{@{}ll@{}}\toprule
				\textbf{Descriptions}& \textbf{Notations}  
				\\ \midrule
				{The traffic demand (split) of BS-$k$} & $\lambda_k^n$ ($i_k^n$)         
				\\ \hdashline
				{Allocated flavors (actual resource utilization) for vDU-$k$/vCU-$k$} &  $x_k^n/y^n_k$ ($\hat{x}_k^n/\hat{y}^n_k$)
				\\ \hdashline
				{Locations of vDU-$k$ and vCU-$k$} &  $z_k^n$, $\zeta_k^n$
				\\ \hdashline
				Maximum computing capacity of FS-$l$ and ES-$m$ & $H_l$, $\hat{H}_m$ 
				\\ \hdashline
				A connecting path of EPC$\rightarrow$ES-$m$, ES-$m$$\rightarrow$FS-$l$, ES-$m$$\rightarrow$RU-$k$, FS-$l$$\rightarrow$RU-$k$ & $p_{0m}$, $p_{ml}$, $p_{mk}$, $p_{lk}$
				\\ \hdashline
				Incurred delay of path $p_{0m}$, $p_{ml}$, $p_{mk}$, $p_{lk}$ & $ d_{p_{0m}}, d_{p_{ml}}, d_{p_{mk}}, d_{p_{lk}}$
				\\ \hdashline
				HLS and LLS delay requirement for split $i$ & $d_i^{H}$, $d_i^{L}$
				\\ \hdashline
				The routing for BS-$k$ & $p \in \mathcal{P}_k$
				\\ \hdashline
				Data flow with split $i$ via routing $p \in \mathcal{P}_k$ (FH, MH , BH) & ($r^{\text{FH}, n}_{p,i}, r^{\text{MH}, n}_{p,i}$, $r^{\text{BH}, n}_{p,i}$)
				\\
				\bottomrule
			\end{tabular}
		\end{small}
	\end{threeparttable}
	\caption{\small Key variables and parameters used in our model.}
	\label{table:notations}
\end{table*}

\begin{figure}[t!]
	\centering
	\includegraphics[width=0.425\textwidth]{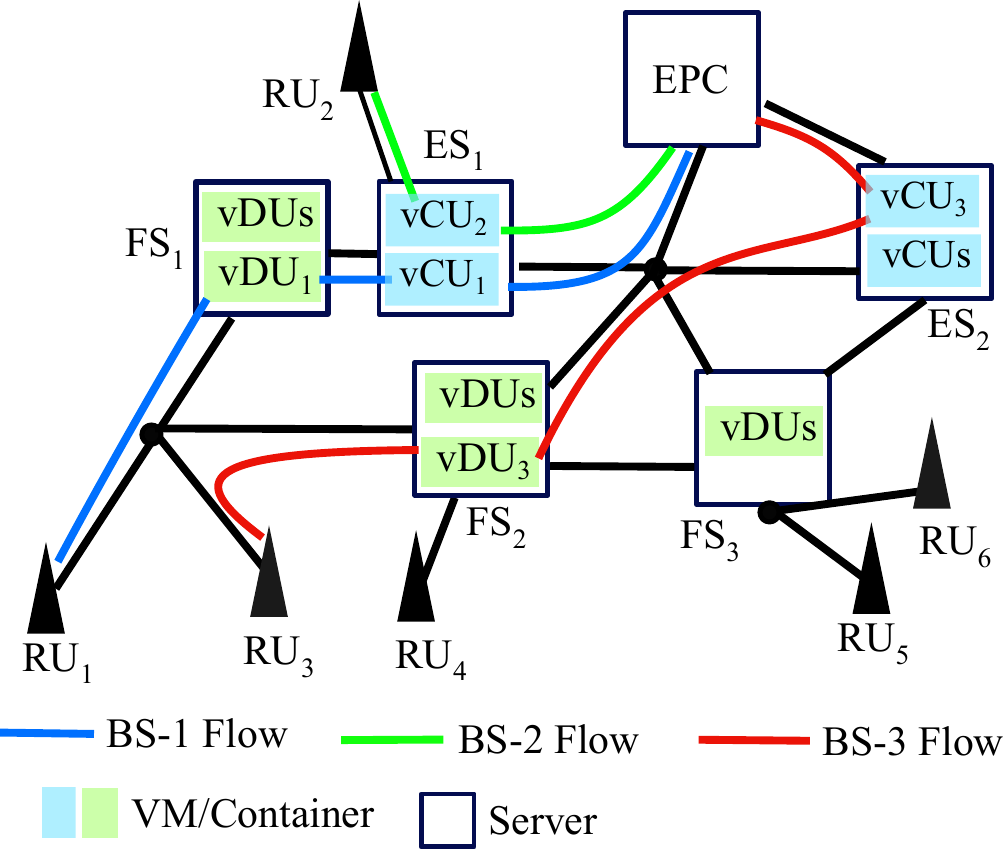}   
	\caption{\small The functions of each BS can be split between the vCU, vDU and RU. For BS-1, vDU-1 and vCU-1 are executed at FS-1 and ES-1, respectively. However, BS-2 implements C-RAN (S4); hence, the integrated vDU/vCU are executed only at ES-1 (e.g., links to and instances at  FSs are not activated). Then, vDU-3 and vCU-3 of BS-3 are hosted at FS-2 and ES-2, respectively. }
	\label{fig:network}
\end{figure}

We consider a vRAN system with $K$ BSs, where the functions of each BS-$k$ can be disaggregated and hosted at vCU-$k$, vDU-$k$ and RU-$k$. The vDUs are executed at far-edge cloud servers (FSs) while the vCUs are at edge cloud servers (ESs)\footnote{FSs are the candidate platforms and locations to execute VM instances of the vDUs. Similarly, ESs are the candidate platforms and locations for the vCUs. We also consider ESs for the candidate platforms to host an integrated vDU/vCU in C-RAN. ESs are typically located at more centralized locations, while FSs are co-located or near the RUs.}. 
%
%
%
%
%
We model a packet-based vRAN as a graph of $G= (\mathcal{V}, \mathcal{E})$, where the set of physical nodes $\mathcal{V}$ includes the subsets: $\mathcal{K} = \{1, ..., K\}$ of RUs, $\mathcal{L} = \{1, ..., L\}$ of FSs, $\mathcal{M} = \{1, ..., M\}$ of ESs, EPC (index 0), and routers. These nodes are connected through a set of links $\mathcal{E}$, where each link $(i,j) \in \mathcal{E}$ has a data transfer capacity $c_{ij}$ (Gbps). 
We denote $\mathcal{P}_k$ as a set of paths connecting EPC to RU-$k$ and consider the data flow for each BS is unsplittable.
We focus on the downlink, but it is not limited and can easily be extended for uplink.
The data flow for each BS will be transferred from EPC to RU-$k$ through a path $p := \{ (0, i_1), (i_1, i_2), ..., (i_k, k) : (i,j) \in \mathcal{E}  \} \in \mathcal{P}_k$. Since this path might pass through FSs and ESs before reaching each RU, let us denote $p_{0m}$, $p_{ml}$, $p_{mk}$, and $p_{lk}$ as a path connecting EPC$\rightarrow$ES-$m$, ES-$m$$\rightarrow$FS-$l$, ES-$m$$\rightarrow$RU-$k$, and FS-$l$$\rightarrow$RU-$k$, respectively.
Based on the selected split, the data flow of each BS-$k$ passes through $p := p_{0m} \cup p_{ml} \cup p_{lk} \in \mathcal{P}_k$ (EPC $\rightarrow$ ES-$m$ $\rightarrow$ FS-$l$ $\rightarrow$ RU-$k$) if activating S1, S2 and S3. Otherwise (e.g., S4/C-RAN), the flow passes through $p := p_{0m} \cup p_{mk} \in \mathcal{P}_k$ (EPC $\rightarrow$ ES-$m$ $\rightarrow$ RU-$k$ without using FSs). 
Each path has a total delay defined as $d_p, d_{p_{0m}}, d_{p_{ml}}, d_{p_{mk}}$ and $d_{p_{lk}}$; and they must respect the delay requirements of the split as described in Table \ref{table:3gpp_split}.
We compute each $p_{0m}$, $p_{ml}$, $p_{mk}$ and $p_{lk}$ with the shortest path method.
%
%
%
Fig. \ref{fig:network} shows an example of our model.
 %

We use the term \textit{flavor}\footnote{This term is carried out from OpenStack (https://www.openstack.org/) to reserve the amount of virtual CPU, memory, and storage capacity for a VM instance. This term is typically used to calculate the billing units to charge the amount of monetary cost. Similar terms are also used in other cloud services such as AWS and Azure. Here, we focus on the CPU resources as they are the most affected performance by the traffic demands.} to define the available choices for allocating the virtualized computing resources.
Let us introduce $\mathcal{X} $ as a set of available flavors for the vDUs and vCUs. Then, we select a flavor $x_k \in \mathcal{X}$ and $y_k \in \mathcal{X}$ that determine the reserved resources for each vDU-$k$ (in FSs) and vCU-$k$ (in ESs).
%
%
%
%
%
%
Each FS-$l$ has physical computing capacity $H_l$, respectively $\hat{H}_m$ for ES-$m$, which bound the aggregate allocated resources (accordingly, the flavors that can be selected) of the vDUs and vCUs for each location. \rev{The key notations used in our model are summarized in Table \ref{table:notations}.}


\subsection{Problem Formulation}

\rev{We model the vRAN operation as a time-slotted system.} Given an incoming sequence of possibly-different traffic demands and resource availability, we aim to design a policy (strategy) of an agent that controls the vRAN configurations at each time slot, which includes the splits of the BSs, flavors and locations of vDUs and vCUs, and the routing for each BS data flow, to minimize the long-term total network operation cost. This sequential decision problem is formulated as MDP, specified \rev{by} a tuple $\{   \mathcal{S}, \mathcal{A}, {P} , r  \}$. At every time slot $n$, the agent observes a state from the state space $s^n \in \mathcal{S}$, then takes an action that selects the vRAN configurations from the action space $a^n \in \mathcal{A}$.
Following each enforced action, the agent expects to receive a reward signal $r(s^n,a^n)$ as feedback from the environment (vRAN system). Since the state may not be stationary, we define ${P} (s^{n+1} | s^n,a^n)$ as the state transition probability that maps a state-action pair at time step $n$ into the distribution of next states. And we take no assumption about it.
%
The formulated problem is also naturally an RL problem, and we describe it as follow.

\subsubsection{Action}  We introduce $i^n := \{i_k^{n} \in \mathcal{I} :  k \in \mathcal{K} \}$ as control variables to select the functional splits that decide which functions of the BSs to be placed at the vDUs and vCUs. The selection of the flavors that allocates the resources for the vDUs and vCUs is determined using control variables $x^n := \{ x_k^{n} \in \mathcal{X} :  k \in \mathcal{K} \}$ and $y^n := \{ {y}_k^{n} \in {\mathcal{X}} :  k \in \mathcal{K} \} $, respectively. We can determine the locations of vDUs over FSs and vCUs over ESs by $z^n := \{ z_k^{n} \in \mathcal{L} :  k \in \mathcal{K} \}$ and $\zeta^n := \{\zeta_k^{n}\in \mathcal{M} :  k \in \mathcal{K} \}$. The routing paths to transferred the data flow of each BS is selected through variables  $p^n := \{ p^n \in \mathcal{P}_k : k \in \mathcal{K} \}$. Since routing variable $p^n \in \mathcal{P}_k$ depends on the placement of the vDU and vCU, we can determine $p := \{ p_{0m} \cup p_{ml} \cup p_{mk} \cup p_{lk} \} \!\in\! \mathcal{P}_k$ directly from $i_k^n, z_k^n$ and $\zeta_k^n$. For instance, if BS-5 with $i_5^n := \text{S1}$ decides $z_5^n := 1$ and $\zeta_5^n := 2$, then the selected path becomes $p := \{ p_{0,2} \cup p_{2,1} \cup \O \cup p_{1,5} \} \!\in\! \mathcal{P}_5$ with the transferred data flow EPC$\rightarrow$ES-2$\rightarrow$FS-1$\rightarrow$RU-5. Therefore, we can treat $p^n \in \mathcal{P}_k$ as part of the environment. Then, we formalize the action at time slot $n$ as:
\begin{align} \label{eq:action_rl}
	{a}^n \!=\! \{i^n, x^n, y^n, z^n, \zeta^n \} \!\in\! \mathcal{A}, \ \mathcal{A} \!:=\! \{\mathcal{I} \!\times\! \mathcal{X}^2 \!\times\! \mathcal{L} \!\times\! \mathcal{M} \}^{|\mathcal{K}|}, 
\end{align}
where this action is taken from the action space $\mathcal{A}$ of a finite set that includes all possible pairs of the reconfiguration control decisions from all the BSs. 

\subsubsection{State} The state observation at each time slot $n$ of the RL problem consists of \textit{(i)} The incoming traffic demands of the BSs $\lambda^{n} := \{ \lambda^{n}_k \in \mathbb{R}_+ : k \in \mathcal{K}  \}$ (Gbps); \textit{(ii)} the previous deployed splits $i^{n-1} := \{i^{n-1}_k \in \mathcal{I} : k \in \mathcal{K} \}$; \textit{(iii)} the previous allocated resources (flavors) for the vDUs $x^{n-1} := \{ x_k^{n-1} \in \mathcal{X} :  k \in \mathcal{K} \}$ and \textit{(iv)} vCUs $y^{n-1} := \{ y_k^{n-1} \in \mathcal{X} :  k \in \mathcal{K} \}$;  and \textit{(v)} the previous deployed locations of each vDU-$k$ over FS $z^{n-1} := \{z_k^{n-1} \in \mathcal{L} : k \in \mathcal{ K}\}$ and \textit{(v)} each vCU-$k$ over ES $\zeta^{n-1} := \{\zeta_k^{n-1} \in \mathcal{M} : k \in \mathcal{K}\}$. It provides \emph{time dynamic} of our variable interests: \textit{(i)} the demand that needs to be served by each BS; \textit{(ii)} the current active splits of the BSs; \textit{(iii)} the availability of resources for each vDU and \textit{(iv)} vCU; and \textit{(v)} the availability to execute each vDU at FS and \textit{(vi)} each vCU at ES. Then, the state observation at time slot $n$ can be denoted:
\begin{align} \label{eq:state_rl}
	{s}^n &:= \{ \lambda^n, i^{n-1}, x^{n-1}, y^{n-1}, z^{n-1}, \zeta^{n-1} \} \in \mathcal{S}, \notag \\
	& \ \mathcal{S} :=  \{ \mathbb{R} \times  \mathcal{I} \times \mathcal{X}^2 \times \mathcal{L} \times \mathcal{{M}} \}^{|\mathcal{K}|}.
\end{align}
The state space $\mathcal{S}$ is semi-continuous because it contains continuous parameters $\lambda_k^n \in \mathbb{R}_+, \forall k \in \mathcal{K}$ from the traffic demands. It is exogenous parameter, i.e., it is not affected by the action, but it provides contextual information about the users' needs. The other points are discrete parameters and provide the network state information, which are highly affected by the deployed configurations from the last action. \rev{This state information is provided as input to the learning agent through the O1 interface. The state can be extended to other relevant key performance measurements; however, the state space of the RL problem also expands.}


%


\subsubsection{Reward \& Policy}Our reward function is calculated from the incurred total network operating cost. The source of monetary costs comes from the computing cost to execute the BS functions, the virtualized resource management costs and the routing cost.

%
%
%
%

The needs of computing cost of each BS-$k$ to host its functions at the RU-$k$, vDU-$k$ (in the FS) and vCU-$k$ (in the ES) are denoted as:
\begin{align} \label{eq:computing_cost}
	f_\text{RU} (\hat{w}_k^{n} ), 
	f_\text{FS} (\hat{x}^{n} ), \text{and }
	f_\text{ES} (\hat{y}_k^{n} ), 
\end{align}
where $f_\text{RU} (.)$, $f_\text{FS} (.)$ and $f_\text{ES} (.)$ are the cost functions to charge the utilized computing processing at the RU\footnote{RUs are the radio hardware units; hence we do not allocate resources for RU. Instead, the computing cost of the RUs is incurred from processing the LP/RF functions, where their processing cost is demand/split dependent.}, FS and ES, respectively. These cost functions translate the actual computing resource utilization of the RUs $\hat{w}^{n} := \{\hat{w}_k^{n} \in \mathbb{R} : k \in \mathcal{K} \}$, vDUs $\hat{x}^{n} := \{\hat{x}_k^{n} \in \mathbb{R} : k \in \mathcal{K} \}$ and vCUs $\hat{y}^{n} := \{\hat{y}_k^{n} \in \mathbb{R} : k \in \mathcal{K} \}$ into monetary units (\$). The actual resource utilization of each RU, vDU and vCU is highly affected by the split and demand at the BS. Hence, we define $\psi : (\lambda_k^{n},i_k^{n}) \!\mapsto\! ( \hat{w}_k^n, \hat{x}_k^n, \hat{y}_k^n )$ as a function to map inputs of the split and traffic demand of the BS into the actual resource utilization at the RU, vDU and vCU. This function represents the actual computing behavior in the vRAN system, and we characterize it through traces from the testbed measurements. 
Further, we consider that cost functions $f_\text{RU} (.)$, $f_\text{FS} (.)$ and $f_\text{ES} (.)$ to be proportional with their input, e.g., $f_\text{FS} (v) := \kappa_\text{RU} v$, $f_\text{FS} (v) := \kappa_\text{FS} v$ and $f_\text{ES} (v ) := \kappa_\text{ES} v$, where $\kappa_\text{RU}$ (\$/unit), $\kappa_\text{FS}$ (\$/unit) and $\kappa_\text{ES}$(\$/unit) are the estimated computing processing fees per core unit capacity at the RUs, FSs and ESs, respectively.

\begin{figure}[t!]
	\centering
	\includegraphics[width=0.42\textwidth]{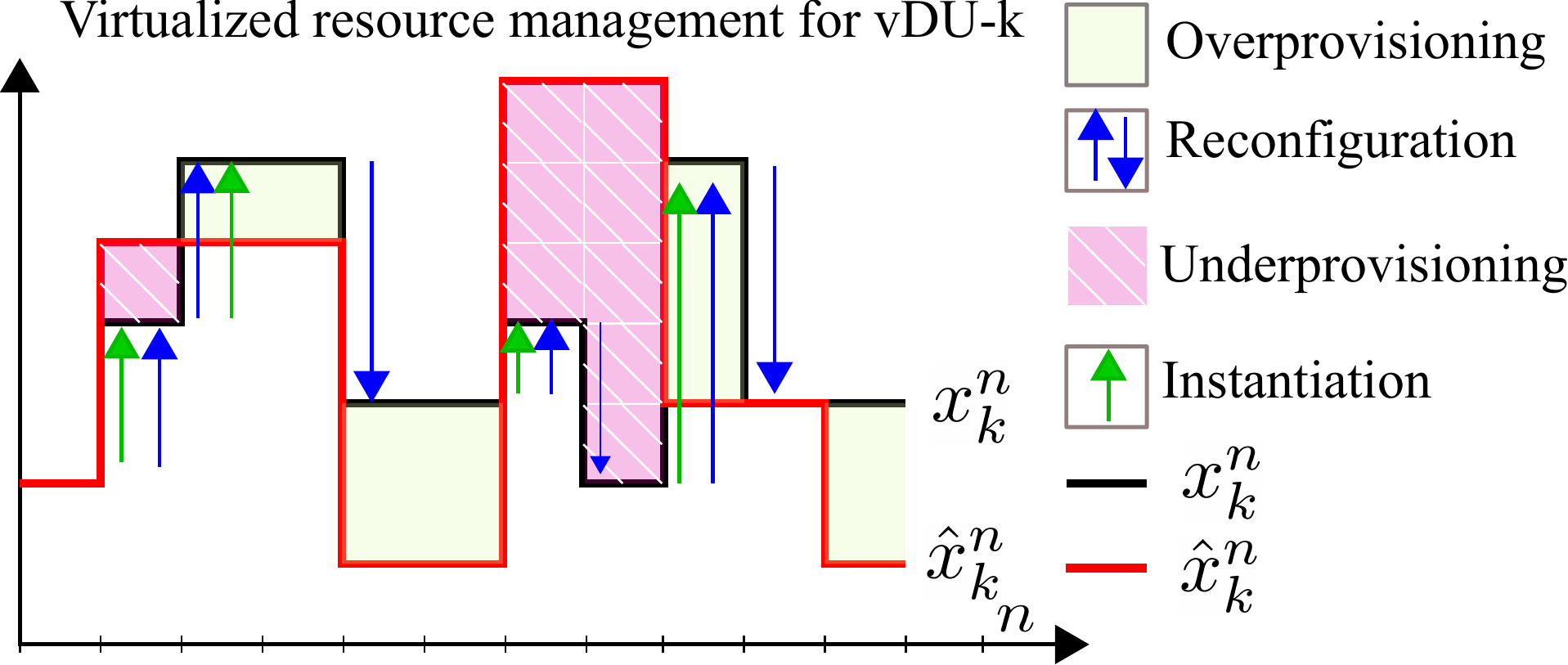}   
	\caption{\small An example of virtualized resource management model for vDU-$k$. }
	\label{fig:orchestration}
	\vspace{-2mm}
\end{figure}

%
In vRANs, the vDUs and vCUs are virtualized on the FSs and ESs, respectively. Therefore, the virtualized resources of the vDUs and vCUs can be dynamically allocated to obtain cost-efficient network operations. However, reconfiguring such resources might \rev{lead} to additional costs.
%
Meanwhile, the allocated resources $x^{n}_k$ and $y^{n}_k$ might differ to the actual resource utilization of $\hat{x}^{n}_k$ and $\hat{y}^{n}_k$, which can create unwanted resource overprovisioning or declined demands. Motivated by resource management in network slicing \cite{aztec}, we propose a cost model capturing such behaviors in vRANs. This model is illustrated in Fig. \ref{fig:orchestration} and described as follows.


%
%

\noindent
\textit{(i) \underline{Overprovisioning}:} If the allocated resources are higher than their actual utilization, the operators pay more expenses and miss the opportunity to share their unused resources for other workloads. Such resources are instantiated and reserved for no purpose, which can be more profitable to be allocated for other workloads (e.g., video analytics) to increase the global system efficiency. This overprovisioning cost at time slot $n$ for BS-$k$ is defined as: 
%
%
\begin{align} \label{eq:overprov_cost}
	f_\text{O}  \big(\max(0,  x_k^{n}  - \hat{x}_k^{n} ) +  \max(0,  y_k^{n}  - \hat{y}_k^{n} ) \big), 
\end{align}
%
where $f_\text{O}(.)$ is a cost function for resource overprovisioning. This function is proportional with the input, e.g. $f_\text{O} (v) \!:=\! \kappa_\text{O} v $, where $\kappa_\text{O}$ is the estimated fee for one unit capacity (\$/unit).

\noindent
\textit{(ii) \underline{Declined service demands:} } The declined demands can occur when there exists an insufficient resource allocation or constraint violation, which triggers service level agreement (SLA) violation and monetary compensation. For instance, the constraint violation can happen when the total allocated resources of the vDUs exceed FS capacity:
\begin{align} \label{eq:constrain_fs}
	f_\text{D}  \Big( \max \big( 0, \sum_{k \in \mathcal{K}} x_k^n \mathbbm{1}_{=l} (z_k^n)   -  H_l \big) \Big), \ \forall l \in \mathcal{L},
\end{align}
the total allocated resources of the vCUs exceed ES capacity:
\begin{align} \label{eq:constrain_es}
	f_\text{D} \Big( \max \big( 0, \sum_{k \in \mathcal{K}} y_k^n \mathbbm{1}_{=m} (\zeta_k^n)   - \hat{H}_m \big) \Big), \ \forall m \in \mathcal{M},
\end{align}
and the incurred delay does not meet the requirement:
\begin{align}
	f_\text{D}  \big(  \max (0, & d_{p_{ml}} - d_i^{{H}}, d_{p_{lk}} - d_i^{{L}}, \notag \\
	& d_{p_{mk}} - 0.25 ) \big), \forall m \in \mathcal{M},  \forall l \in \mathcal{L},
\end{align}
where $d_i^{{H}}$ and $d_i^{{L}}$ are the delay requirement of split $i$ for the HLS and LLS, respectively, as defined in Table \ref{table:3gpp_split}. In addition to the constraint violation, an insufficient allocation for each vDU and vCU can cause declined service demands, and we define this as:
\begin{align} \label{eq:underprov_cost}
	f_\text{D} \big(
	\max ( 0, \hat{x}_k^{n} - x_k^{n}, \hat{y}_k^{n} - y_k^{n} ) \big).
\end{align}
 The function $f_\text{D}(.)$ captures the monetary compensation that the operators have to pay for violating the SLA. This function is assumed to be proportional with the input, e.g. $f_\text{D} (v) \!:=\! \kappa_\text{D} v $, where $\kappa_\text{D}$ is the estimated fee for declined demands in one unit capacity (\$/unit).

\noindent
\textit{(iii) \underline{Instantiation and Reconfiguration:}} 
The operators may decide to instantiate new resources or reconfigure their network settings to reduce resource overprovisioning and declined demands and adapt to the varying traffic demands and resource availability.  However, instantiating and reconfiguring such resources (e.g., VMs) induce capital expenses, and we define it as:
%
%
%
%
\begin{align} \label{eq:insta_cost}
	&f_\text{I} \big( \max ( 0,  x_k^{n} - x_k^{n-1}) +  \max( 0,  y_k^{n} - y_k^{n-1}) \big),
\end{align}
\begin{align} \label{eq:reconfigure_cost}
	f_{\text{R}} \Big( \big( |x_k^{n} &- x_k^{n-1}| +  |y_k^{n} - y_k^{n-1}| \big) + \notag \\
	\ \ &\big( x_k^{n} \mathbbm{1}_{\neq z_k^{n-1}} (z_k^n) + y_k^{n} \mathbbm{1}_{\neq \zeta_k^{n-1}} (\zeta_k^n) \big) \Big),
\end{align}
%
%
where $f_\text{I}(.)$ and $f_\text{R}(.)$ are the cost functions for resource instantiation and reconfiguration.
Eq. \eqref{eq:insta_cost} captures the amount of instantiating additional resources for the vDU and vCU, which might arise due to migrating additional resources to serve the vRAN workload, and this results in indirect overhead expenses such as the increase of power consumption \cite{aztec}.
%
%
Then, the first term in \eqref{eq:reconfigure_cost} captures the reconfiguration cost initiated from migration activities for altering the splits and  flavors (resizing resources). 
\rev{Such activities raise overhead costs from the migrated resources, measured from the difference between the current and the previous resources \cite{aztec, adaptive_vran_alba}.
For instance, altering the splits requires creating new BS functions while maintaining the old migrated functions to keep active \cite{adaptive_vran_alba}. Resizing the VMs' resources also initiates a price of management delay \cite{5Gcoral} as it needs time for migrating (and bootstrapping) the computing resources, load balancing and steering the network load}\footnote{We have calculated the incurred time for resizing a VM instance in CSC cPouta (https://www.csc.fi/) cloud computing platform, and it takes around 25 seconds. Modern software architecture such as Kubernetes also requires several seconds to executing new pods \cite{aztec}.}.
The second term in \eqref{eq:reconfigure_cost} captures the reconfiguration cost for migrating the vDU and vCU instances to other FS and ES locations.  In this case, the whole resources of vDU and vCU instances are affected, and the attached routing paths need to be recomputed with the new FS and ES locations.
In our evaluation, $f_\text{I}(.)$ and $f_\text{R}(.)$ are proportional to the input, e.g., $f_\text{I}(v) \!:=\! \kappa_\text{I} v$ and $f_\text{R}(v) \!:=\! \kappa_\text{R} v$, where $\kappa_\text{I}$ (\$/unit) is the estimated cost for resource instantiation and $\kappa_\text{R}$ (\$/unit) is for reconfiguration. If reconfiguring the system \rev{does} not incur any overhead cost, we can set $\kappa_\text{R} = 0$, otherwise $\kappa_\text{R} > 0$.


O-RAN has encouraged adopting an open interface between the vCUs, vDUs and RUs \cite{oran_architecture}, resulting in sharing the xHaul links among the BSs. 
In addition, S1, S2, S3 and S4 generate different data loads depending on the selected split as seen in Table \ref{table:3gpp_split}. Hence, the cost for reserving bandwidth and routing the data flow through the xHaul links are also different. 
The routing cost for each BS-$k$ can be denoted as: 
\begin{align}\label{eq:cost_xhaul}
	f_\text{H} \Big( & \sum_{p \in \mathcal{P}_k} \big( r^{\text{FH}, n}_{p,i} \sum_{l \in \mathcal{L}} \mathbbm{1}_{=z_k^n} (l) \notag \\ 
	& \ + r^{\text{MH}, n}_{p,i} \sum_{m \in \mathcal{M}} \mathbbm{1}_{=\zeta_k^n} (m) +r^{\text{BH}, n}_{p,i} \big) \Big),
\end{align}
where $r^{\text{FH}, n}_{p,i}$, $r^{\text{MH}, n}_{p,i}$, $r^{\text{BH}, n}_{p,i}$ are the incurred data loads over FH, MH and BH at time slot $n$ from using path $p $, serving traffic demand $\lambda$, and deploying split-$i$. The indicator $\mathbbm{1}_{=z_k^n} (l)$ activates if vDU-$k$ is placed at FS-$l$ and $\mathbbm{1}_{=\zeta_k^n} (m)$ activates if vCU-$k$ is hosted at ES-$m$. 
%
%
Then, $f_\text{H}(.)$ is the cost function for bandwidth reservation to transfer data load through the xHaul links, and this cost function is proportional with the input, e.g., $f_\text{H}(v) := \kappa_\text{H}^p v$, where $\kappa_\text{H}^p$ (\$/Gbps/Km) is the estimated fee for reserving bandwidth for path $p$ per Gbps/Km.

%
Let suppose $J^n ({a}^n, {s}^n)  := \sum_{k \in \mathcal{K}} f_{\text{RU}} (.) + f_{\text{FS}} (.) + f_{\text{ES}} (.) + f_{\text{O}} (.) + f_{\text{D}} (.) + f_{\text{I}} (.) + f_{\text{R}} (.) + f_{\text{H}} (.) $ is the total operation cost for all the BSs accounted from \eqref{eq:computing_cost}-\eqref{eq:cost_xhaul}. Then, we define the reward\footnote{\rev{Our study focuses on network operation cost minimization, but our framework can be extended to other or multiple objectives, such as maximizing the vRAN performance (e.g., centralization degree). In this case, we can use weighting parameters that determine the relative importance between the objectives (e.g., cost and performance).}}:
\begin{align} \label{eq:reward_function_rl}
	r({a}^n, {s}^n) := - J^n({a}^n, {s}^n).
\end{align}

%
%
%
 %
\rev{Then, our aim is to} design \textit{an optimal policy} that maps the input state observation into action $\pi^*({s}) \!:\! \mathcal{S} \!\mapsto\! \mathcal{A}$ , which minimizes the long-term total operation cost over period of time. Such a policy can be formulated through maximizing the long-term reward: 
%
\begin{align}\label{eq:objective_rl}
	\pi_{*}  := \arg\max \mathbb{E}_\pi \big[ \sum_{\tau=0}^{N} \gamma^\tau r^{\tau + n} | \pi \big],
\end{align}
where $\mathbb{E} \big[ \sum_{\tau=0}^{N} \gamma^\tau r^{\tau + n}\big]$ is the expected long-term accumulated reward starting at time slot $\tau$. 
The discount factor $\gamma$ is strictly set to $\gamma =1 $ during the online operation, corresponding to a non-discounted reward that represents the actual cost; otherwise, $\gamma \! \in \! (0,1]$.

\vspace{-1mm}
\subsection{Trade-offs}

The above problem is intricate for many reasons.
 We discuss the trade-offs and non-triviality that arise as follow.

\textbf{(i)} From S1 to S4, the operators can gain a lower computational cost and high-performance operations through function centralization. However, it also has a tighter constraint requirement and induces a higher transferred data load through the xHaul links. A higher data load means a more expensive routing cost. In addition to the splits, the required resources for the vDUs and vCUs are highly affected by traffic demands and resource availability, which might change absurdly. These also affect the placement of the vDUs and vCUs over FSs and ESs. The association and routing paths are also different for each placement location.

\textbf{(ii)} Using a static policy and finding the best configurations by foreseeing the future peak traffic may reduce the overhead costs due to reconfiguration activities. However, it might produce significant resource overprovisioning. Such unused resources can be profitable if the operators can efficiently manage and share with other workloads. Predicting the future peak traffic might also be inaccurate, which might not result in the best configurations.

\textbf{(iii)} By dynamically reconfiguring the vRAN settings at every time slot, the operators can obtain the best configurations at a time; hence, the risks of resource overprovisioning and declined demands can be reduced. However, every reconfiguration activity produces overhead costs, which may lead to costly long-term network operations. Moreover, the reconfiguration decisions are made before the actual traffic demand is observed; therefore, finding the optimal decisions at every time slot is challenging and might be unfeasible in practice.

\textbf{(iv)} The reconfiguration decisions in our vRAN system are highly affected by the traffic demands and resource utilization. However, their relations are complex, depending on many factors such as traffic demand, computing platform, radio scheduler, etc, which also hinder general assumptions (e.g., linear) to model the computing resource's behavior, rendering traditional control policies inefficient for our vRAN reconfiguration problem. 


\textbf{(v)} Points (i)-(iv) emphasize the need for intelligent reconfiguration decisions with minimal assumptions about the underlying system. A deep RL paradigm can be suitable to handle such challenges. However, the formulated RL problem has a huge state space and multi-dimensional action space because the vRAN system consists of multiple BSs sharing the same network resources with highly coupled configuration decisions. These challenges make conventional deep RL discrete action space algorithms such as deep Q learning inefficient. 

Given the formulated RL problem and trade-offs above, we present how to design the solution that solves the problem efficiently in the next section.


%

%% file: Algorithms.tex
\section{LARV Learning Algorithm} \label{sec:algo}
LARV leverages a model-free RL paradigm, which considers the vRAN system as a black-box environment and does not take any assumption about the system state and state transition probability distribution. 
However, finding the optimal policy of the agent is non-trivial as the formulated RL problem has the semi-continuous state space and the multi-dimensional action space, which make the state-action space extremely large. The large state space can be addressed using D3QN \cite{dueling_dqn}, where this approach is also naturally designed for discrete action. However, we need to tackle the issue of the multi-dimensional action space, which makes the number of estimated actions grow combinatorially with the number of BSs and configuration decisions. In order to address this curse dimensionality, we incorporate action branching \cite{bdq} with D3QN to compress the number of estimated actions. Through this approach, the multi-dimensions of the action can be distributed across individual network branches while maintaining a shared decision module among them to encode a latent representation of the input state and enable coordination among the branches. In contrast to traditional discrete-action deep RL algorithms, this action decomposition method exhibits a linear growth of the total network outputs with increasing action dimensionality.

\vspace{-1mm}
\subsection{D3QN to Address the Large State Space}
%

%

The objective of our RL agent is to learn the optimal policy $\pi_{*}$ defined in \eqref{eq:objective_rl}. As the problem has a large state space and the expected output is a discrete action, we can utilize an off-policy RL algorithm by using D3QN to approximate the action-value function (Q-function) and Double Q-learning for the learning step.

We define the optimal action-value function $Q^*(s, a)$ as the maximum expected reward for observing certain sequences $s$ after following some policies $\pi$ and taking some actions $a$ as: $Q^*(s, a):= \max_\pi \mathbb{E}[\sum_\tau^\infty \gamma r^{\tau+n} | s^n = s, a^n = a]$. If we know the optimal value $Q^*(s', a')$ of the sequence at the next time slot $s'$ for all possible actions $a'$, we can identify the optimal policy $\pi^{*}$, which is to select action $a'$ that maximizes the expected value $r + \gamma Q^*(s', a')$: $Q^*(s, a) := \mathbb{E}_{s \sim \mathcal{E}} [r + \gamma \max_{a'} Q^*(s',a') | s', a']$. In the value iteration method, the action-value function can converge to the optimality when the iteration number reaches near infinity; however, it is impractical.
 Therefore, a function approximator such as a neural network can be applied to estimate the action-value function. The estimated action-value function parameterized by a neural network (Q-network) with weights $\theta$ is denoted as: $Q(s,a;\theta) \approx Q(s,a)$. Then, the Q-network is trained by minimization of a loss function: 
\begin{align}\label{eq:loss_dqn}
	L(\theta) := \mathbb{E}_{s, a,r, s' \sim \mathcal{D}} \big[ u - Q(s, a; \theta) \big],
\end{align}
where the transition $\{s, a,r, s' \}$ is collected through random sampling (minibatches) from stored experience data $\mathcal{D}$, and $u$ is the Temporal Difference (TD) target. In DQN \cite{dqn_mnih1}, the TD target is computed by:
\begin{align}\label{eq:target_dqn}
	u^{\text{DQN}} := \mathbb{E}_{s' \sim \mathcal{S}} [r + \gamma \max \tilde{Q}(s', a'; \tilde{\theta})],
\end{align}
where $\tilde{Q}(s', a'; \tilde{\theta})$ is the target network parameterized by weights $\tilde{\theta}$. The design of TD-target in \eqref{eq:target_dqn} often causes an overestimate to the actual action-value. Thus, we apply Double DQN (DDQN) \cite{ddqn} to overcome this issue by modifying the TD target into: 
\begin{align}\label{eq:target_ddqn}
	u^{\text{DDQN}} := \mathbb{E}_{s' \sim \mathcal{S}} [r + \gamma \tilde{Q}(s', \underset{a'}{\arg\max} Q(s', a'; \theta); \tilde{\theta})].
\end{align} 
When the RL problem has a large action space, such as in our vRAN problem, it might not require estimating the value for certain states, i.e., avoiding unnecessary estimation of redundant and low-value actions. Thus, we apply the Dueling architecture \cite{dueling_dqn} to DDQN (called D3QN) by separating the Q-network into two streams of state-value and advantage, which are then combined through an aggregating layer to produce an estimate of the action-value function.
Lets denote $V(s; \theta)$ and $A(s, a; \theta)$ as the estimated state-value function and advantage function, respectively; then, the action-value function at the output layer can be computed as: 
\begin{align}\label{eq:dueling_ddqn}
	{Q}(s, a; \theta) := V(s; \theta) + A(s, a; \theta) - \frac{1}{|\mathcal{A}|} \sum_{a'} A(s, a'; \theta).
\end{align} 
By explicitly separating the Q network into two estimators, D3QN can learn which states are valuable without requiring to learn the impact of every action for each state. Hence, it can effectively achieve a high-quality policy for a large state space. However, in addition to a large state space, our vRAN problem produces a multi-dimensional discrete action space. It drives the number of estimated Q values in \eqref{eq:dueling_ddqn} to grow combinatorially with the number of configuration decisions and BSs. Next, we present how we incorporate an action branching architecture with D3QN to compress the number of estimated Q values in our vRAN problem.

\vspace{-2mm}
\subsection{Action Compression Using Action Branching}
Let us define $\mathcal{C}_k := \{i_k, x_k, y_k, z_k, \zeta_k\}$ as a set that includes all the reconfiguration control variables of BS-$k$. Then, we denote the sub-action $a_{kc}, \forall c \in \mathcal{C}_k, \forall k \in \mathcal{K},$ to represent the $c$-th reconfiguration control variables of BS-$k$, i.e., $a_{11} := i_1, a_{12} := x_1, ..., a_{KC_K} := \zeta_K$; and $C_k := |\mathcal{C}_k|, \forall k \in \mathcal{K}$. Hence, we can rewrite the action in \eqref{eq:action_rl} by $a := \{a_{kc} : c \in \mathcal{C}_k, k \in \mathcal{K}\}$. Each of sub-actions also takes values from a finite set of the sub-action space $\mathcal{A}_{kc} \subseteq \mathcal{A}$ that describes the $c$-th reconfiguration control space of BS-$k$, i.e., $\mathcal{A}_{k1} := \mathcal{I}, \mathcal{A}_{k2} := \mathcal{X}, ..., \mathcal{A}_{kC_K} := \mathcal{M}, \forall k \in \mathcal{K}$.
\rev{As the RL agent controls $K$ BSs, and each BS has $C_k$ sub-actions; then, the number of Q-values to be estimated turn to $\prod_{k=1}^{K} \prod_{c=1}^{C_k} |\mathcal{A}_{kc}|$. By incorporating action branching, the number of Q-values to be estimated can be compressed to $\sum_{k = 1}^K \sum_{c = 1}^{C_k} |\mathcal{A}_{kc}|$. The initial action branching in \cite{bdq} has successfully tackled problems with the discretized continuous action space. However, its performance is still not validated in the problem where the action space is naturally multi-dimensional. Moreover, it assumes that all of the sub-action spaces have the same dimensional size, i.e., $|\mathcal{A}_{11}| = |\mathcal{A}_{12}| =... =|\mathcal{A}_{KC_K}|$.} Hence, we can not directly utilize it as the size of the sub-action space of the reconfiguration control variables in our vRAN problem varies. We adopt the action branching paradigm suited to our problem and describe it as follows.

We use the common state $s$ defined in \eqref{eq:state_rl} and common state-value $V(s)$. The value of sub-action $a_{kc}$ at common state $s$ with the corresponding sub-action advantage $A_{kc} (s, a_{kc})$ becomes:  
\begin{align} \label{eq:q_bdq}
	Q_{kc} (s, a_{kc}) &:= V(s) + \big( A_{kc}(s, a_{kc})
	\notag \\ 
	& \ - \frac{1}{|\mathcal{A}_{kc}|} \sum_{a_{kc}' \in \mathcal{A}_{kc}} A_{kc}(s, a_{kc}') \big).
\end{align}
Then, the TD target is set similar to \eqref{eq:target_ddqn} to avoid maximization bias, except it uses an average of all the dimensions of the sub-actions as follows:
\begin{align}\label{eq:target_bdq}
	u \!:=\! r \!+\! \gamma \frac{1}{K} \sum_{k=1}^K \frac{1}{C_k} \sum_{c=1}^{C_k} \! \tilde{Q}_{kc} \big( s', \underset{a_{kc'} \in \mathcal{A}_{kc}}{\arg\max} \ Q_{kc}(s', a_{kc}') \big),
\end{align}
where $\tilde{Q}_{kc}$ is the target network. Then, the loss function can be computed as:
\begin{align}\label{eq:loss_bdq}
	L(\theta) \! := \!  \mathbb{E}_{s, a, r, s' \sim \mathcal{D}} \big[ \frac{1}{K} \sum_{k = 1}^K \frac{1}{C_k} \sum_{c=1}^{C_k} [u_{kc} \!-\! Q_{kc}(s, a_{kc}; \theta) ] \big].
\end{align}
The action $a$ to be taken for all the BSs is selected based on $\epsilon$-greedy, where the agent chooses a random action with probability $\epsilon$ or compute:
\begin{align} \label{eq:action_bdq}
	a \!:=\! \big[ \underset{a_{k1}'}{\arg\max  Q_{k1} (s, a_{k1'}) }, ...,  \underset{a_{KC}'}{\arg\max  Q_{KC_K} (s, a_{KC_K'}) } \big]
\end{align}
with probability $1- \epsilon$.

\vspace{-2mm}
\subsection{Neural Network Architecture and Learning Algorithm}
\begin{figure*}[t!] 
	\centering
	\includegraphics[width=0.6\textwidth]{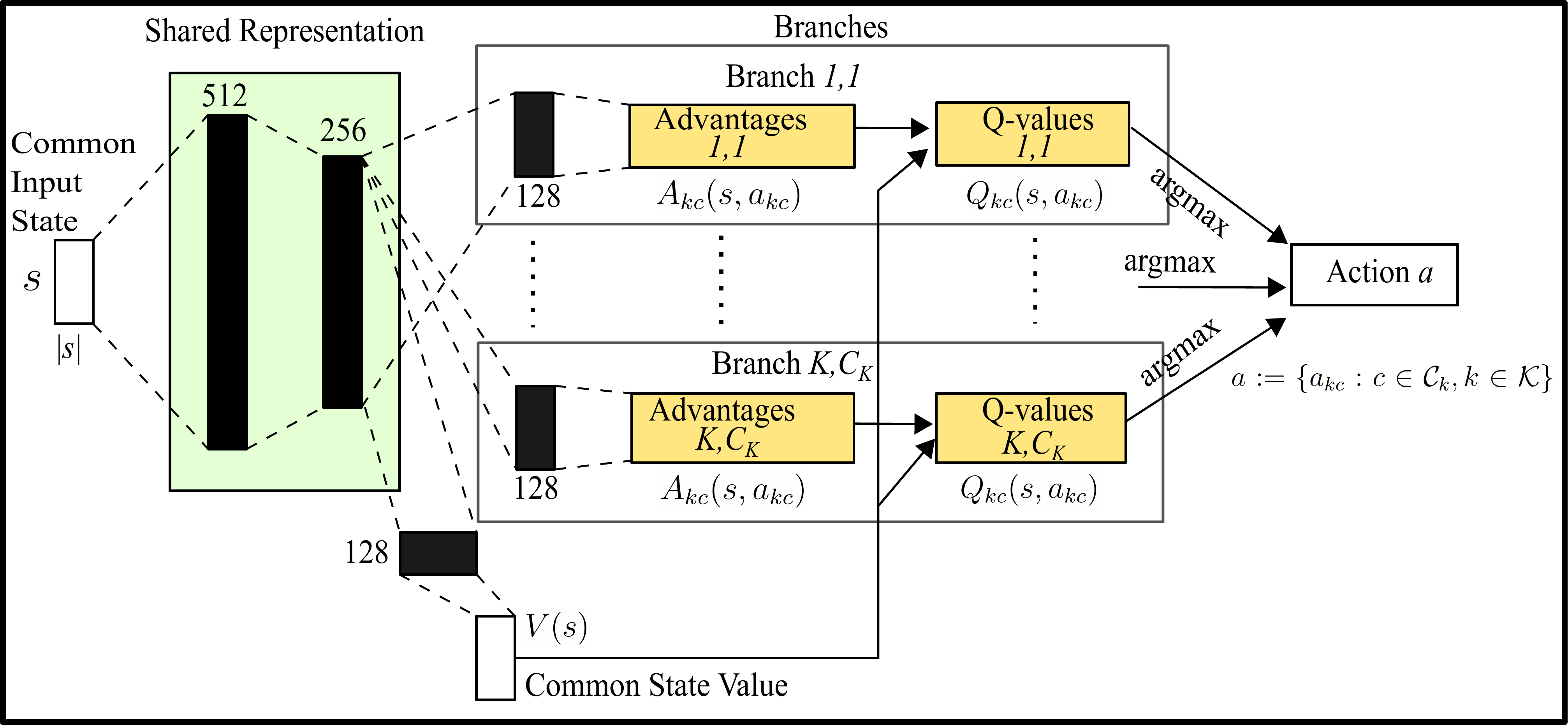}   
	\caption{\small{The Q-network architecture built following D3QN with action branching.}}
	\label{fig:bdq}
	\vspace{-3mm}
\end{figure*}
%


Fig. \ref{fig:bdq} illustrates the Q-network architecture of branching D3QN $Q_\theta$, parameterized by weights $\theta$ and applied in LARV. This network is constructed from an input layer, a shared representation segment comprising hidden layers, a state value network, and neural network branches. The input layer (Linear layer with ReLU activation) receives the common state observation $s$ and has the size of $|s|$. The shared representation segment is built from two fully connected Linear layers with ReLU activation, connected to neural network branches and state value function network. We use a Linear layer for the common state value network. Then, the neural network branches have a total of $\sum_{k=1}^K C_k$ branches corresponding to the number of control decision variables (sub-actions). Each branch aims to produce the sub-action value $Q_{kc} (s, a_{kc})$ by taking consideration of the common state value $V(s)$ and sub-action advantages $A_{kc}(s, a_{kc})$ as described in \eqref{eq:q_bdq}. Each branch has an output layer (\rev{an aggregation layer from the state value and sub-action advantages}) with the size of $|\mathcal{A}_{kc}|$.

\begin{algorithm}[t!]  
	\caption{LARV Learning Algorithm}
	\label{alg:train}
	\SetAlgoLined
	\DontPrintSemicolon
	\textbf{Initialize:} Replay memory $\mathcal{D}$ with a fixed buffer size, Q-network $Q_{\theta}$ (Fig. \ref{fig:bdq}) with random or pretraining weights $\theta$. \\
	Clone Q-network $Q_{\theta}$ to target network $\tilde{Q}_{\tilde{\theta}}$ with weights $\tilde{\theta} \leftarrow \theta$.  \\	
	\For{Each episode $e=1.., E$}{
		Reset state of all the BSs ${s}^1 := \{ \lambda^1, i^{0}, x^{0}, y^{0}, z^{0}, \zeta^{0}\}$. \\
		\For{Each time slot $n=1...,N$}{
			Select an action $a^n := \{a^n_{kc} : c \in \mathcal{C}_k, k \in \mathcal{K}\}$ randomly with probability $\epsilon$, otherwise compute $a^n$ by using \eqref{eq:action_bdq}. \\
			Determine the routing $p \in \mathcal{P}_k, \forall k \in \mathcal{K}$ using $i^n, z^n$ and $\zeta^n$ obtained from $a^n$.\\
			Enforce $a^n$ and $p \in \mathcal{P}_k, \forall k \in \mathcal{K}$ to all the BSs and compute the total cost $J^n$. \\
			Collect the reward $r^n$ based on \eqref{eq:reward_function_rl}. \\
			Set $s^{n+1} \leftarrow s^n$ with the current observation. \\
			Store the experience $\mathcal{D} \leftarrow \big\{s^n, a^n, r^n, s^{n+1}  \big\}$. \\
			Sample minibatch of experiences from $\mathcal{D}$. \\
			Compute TD target $u$ using \eqref{eq:target_bdq} if not done, otherwise $u := r^n$. \\
			Perform a gradient descent method to the loss function $L(\theta)$ in \eqref{eq:loss_bdq} w.r.t $\theta$.
			\\
			Update target network \rev{ $\tilde{Q}_{\tilde{\theta}} \leftarrow Q_{\theta}$} every $\hat{n}$ steps.
		}
	}
	\vspace*{-1mm}
\end{algorithm}

Further, we summarize the learning process of LARV in Algorithm \ref{alg:train}. Firstly, the replay buffer memory $\mathcal{D}$ and the Q-network $Q_\theta$ (Fig. \ref{fig:bdq}) are initialized, where the Q-network initialization can be from random or pretrained weights \textbf{(Step 1)}. 
Then, the weights of the Q-network $Q_\theta$ are copied to the target network $\tilde{Q}_{\tilde{\theta}}$ \textbf{(Step 2)}. 
At the beginning of each episode (or trial during the training), the state observation $s^1$ is reset with initial values, where these values are assigned from $ \lambda_{}^1 := \{\lambda_{k}^1 \in \mathbb{R}_+ : k \in \mathcal{K} \}$,  $ i^0 := \{ i_k^0 = \text{S1} : k \in \mathcal{K} \}$ , $x^{0} := \{x^{0}_k = \max(\mathcal{X}) : k \in \mathcal{K} \}$, $y^{0} := \{y^{0}_k = \max(\mathcal{X}) : k \in \mathcal{K} \}$ , $z^{0} := \{z^{0}_k = \text{random}(\mathcal{X}) : k \in \mathcal{K} \}$ and $\zeta^{0} := \{\zeta^{0}_k = \text{random}(\mathcal{X}) : k \in \mathcal{K} \}$ \textbf{(Step 4)}. 
Then, at every time slot $n$, given the state observation $s^n$, an action $a^n := \{a^n_{kc} : c \in \mathcal{C}_k, k \in \mathcal{K}\}$ is selected randomly with probability $\epsilon$, otherwise it is computed using \eqref{eq:action_bdq} \textbf{(Step 6)}. 
Then, the routing $p := p_{0m} \cup p_{ml} \cup p_{lk} \in \mathcal{P}_k : k \in \mathcal{K}$ can be selected through $i^n, z^n$, and $ \zeta^n$ obtained from the selected action since these variables determine the hosting servers for the vDUs and vCUs, and hence the destination server for each data flow \textbf{(Step 7)}. After all the control variables are determined, they are enforced to all the BSs as the vRAN configurations at time slot $n$. As a result of the deployed configurations, LARV expects to receive the total operation cost $J(a^n, s^n)$ \textbf{(Step 8)}. Based on this cost, the reward $r(a^n, s^n)$ signal at time $n$ can be computed by \eqref{eq:reward_function_rl} \textbf{(Step 9)}. The state is updated with the current observation $s^{n+1} \leftarrow s^n$  \textbf{(Step 10)}. Then, the agent's experience is stored in replay memory $\mathcal{D} \leftarrow \big\{s^n, a^n, r^n, s^{n+1}  \big\}$ \textbf{(Step 11)} and the memory $\mathcal{D}$ is sampled randomly \textbf{(Step 12)}. Further, the TD target of branching D3QN is computed with \eqref{eq:target_bdq}. Once the TD-target is obtained, we can proceed to calculate the loss function $L(\theta)$ using \eqref{eq:loss_bdq} \textbf{(Step 13)}. The goal of this learning process is to minimize this loss function with regards to weights $\theta$, and we rely on Adam optimizer \cite{adam_optim} to perform stochastic gradient descent. Mostly, the target network is frozen, but it is updated every $\hat{n}$ by using the Q-network weights \textbf{(Step 15)}.


%% file: Simulations.tex
\section{Results and Discussion} \label{sec:simulation}
%
In this section, we perform trace-driven simulations using real traces collected from our testbed to evaluate the performance of LARV under various scenarios during the training process and online operation.

\vspace{-2mm}
\subsection{Experimental Setup} \label{subsection:experimental_setup}
\rev{We built a bespoke testbed to collect measurements used to evaluate LARV under realistic conditions. We utilize the software-based srsRAN \cite{srslte}, where each entity is virtualized using container-based virtualization from Docker. The radio interfaces of the BS (e.g., RU) and user are emulated via ZMQ. The srsENB acts as a BBU of the BS. To deal with functional split, we use prior studies that divide the computing consumptions of LP, HP, LM, HM, LR, HR, and PD functions to yield 48\%, 17\%, 7\%, 7\%, 0.5\%, 0.5\%, 10\%, 10\% of the total BBU, respectively, cf. \cite{andres_fluidran_joint, placeran}. }
We deploy the virtualized entities in Platform A (CSC cPouta hpc.5.16core with max. 16 vCPU) and Platform B (PC AMD Ryzen 7 PRO 4750U with max. 16 CPU threads). We use these computing specifications for \textit{Reference Core (RC)}, i.e., 1 RC translates to 1 CPU thread and 1 vCPU. The virtualized resource of each container can be controlled through \textit{--cpus}, which allows us to set a capacity limit and isolate each container resource. We set an initial resource reservation for srsENB with 10 RCs. In our measurements, the trafﬁc demand follows a Poisson-generated user datagram protocol with a peak data rate is 36.6 Mbps (SISO 10 MHz LTE). 

%
%
%

In our simulations, the trafﬁc demands follow the Milan network datasets from Telecom Italia \cite{milan_datasets}, \rev{where each time slot has 10 minutes time interval. This interval is also aligned with the capabilities of current Virtual Infrastructure Managers (VIMs). Moreover, LARV selects an action from the incoming state information (e.g., by passing forward through the Q network) at each time slot, and it can be performed within a second in our test, which is suitable for real-time operation.} 
\rev{The Milan datasets consist of mixed traffic, including calls, sms, and the internet. We filtered the datasets and utilized internet traffic (mobile broadband). Although it was recorded in 2013 (dominated by 4G traffic), it is still relevant for 5G network evaluation since it captures users' demand behavior comprehensively (e.g., the day, night, weekend, city center, etc.).}
Considering the limitations of our testbed and the difficulty in capturing the computing behavior of the Milan traffic in a tractable model, we utilize a deep neural network\footnote{It is constructed from an input, an output and three hidden layers with the sizes of $128$, $64$ and $16$. We use Adam optimizer \cite{adam_optim} with learning rate is set to $5\!\times\! 10^{-5}$, mini-batch with the size of 128 and MSE loss function, then train it with 200 epochs.} to map the Milan traffic demands into the actual resource utilization, trained using our collected measurements. 

\rev{\begin{figure}[t!]
	\centering
	\begin{subfigure}[t]{.225\textwidth}
		\includegraphics[width=\textwidth]{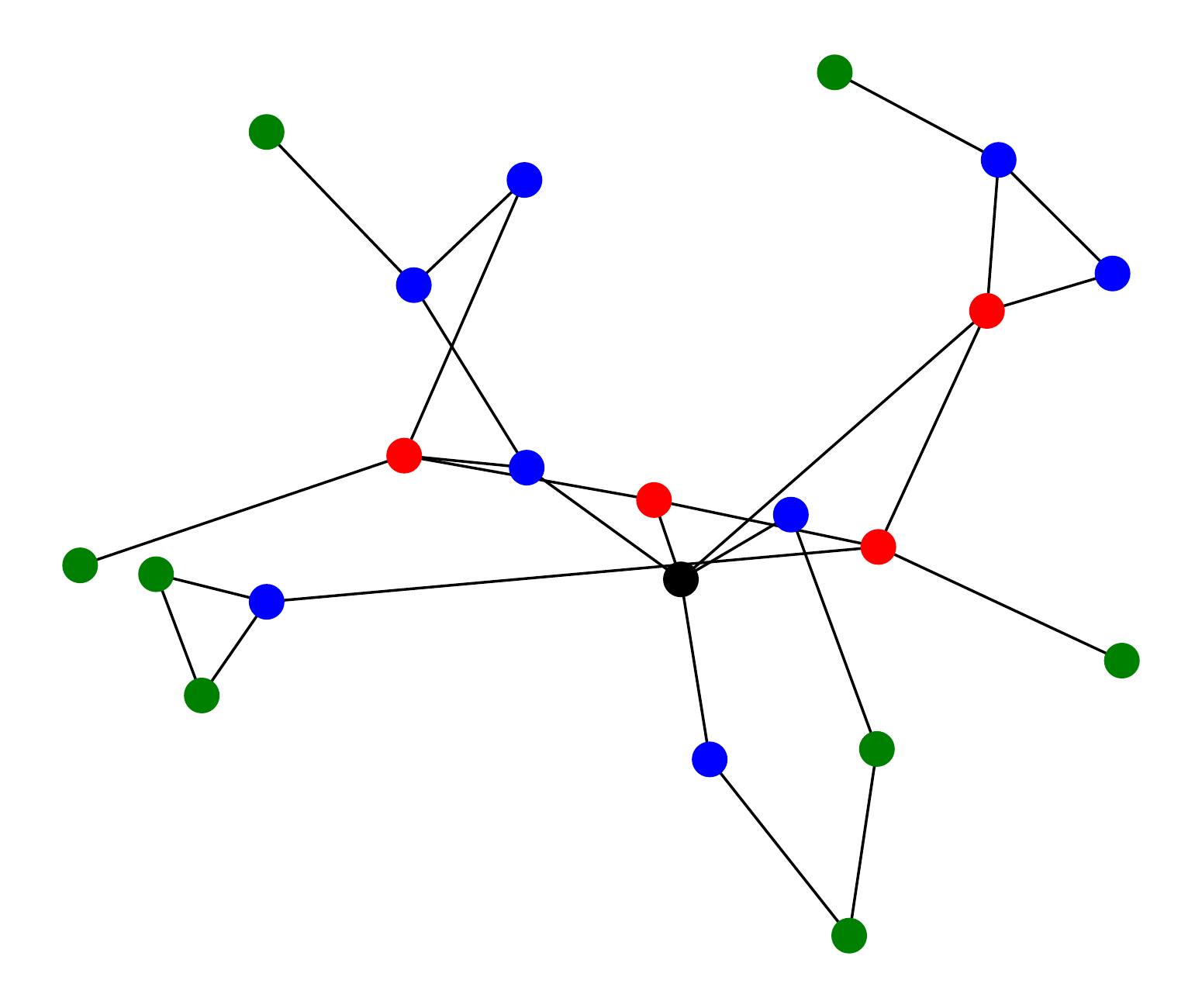}
		\caption{}
		\label{fig:topo_milan}
	\end{subfigure} 
	\begin{subfigure}[t]{.225\textwidth}
		\includegraphics[width=\textwidth]{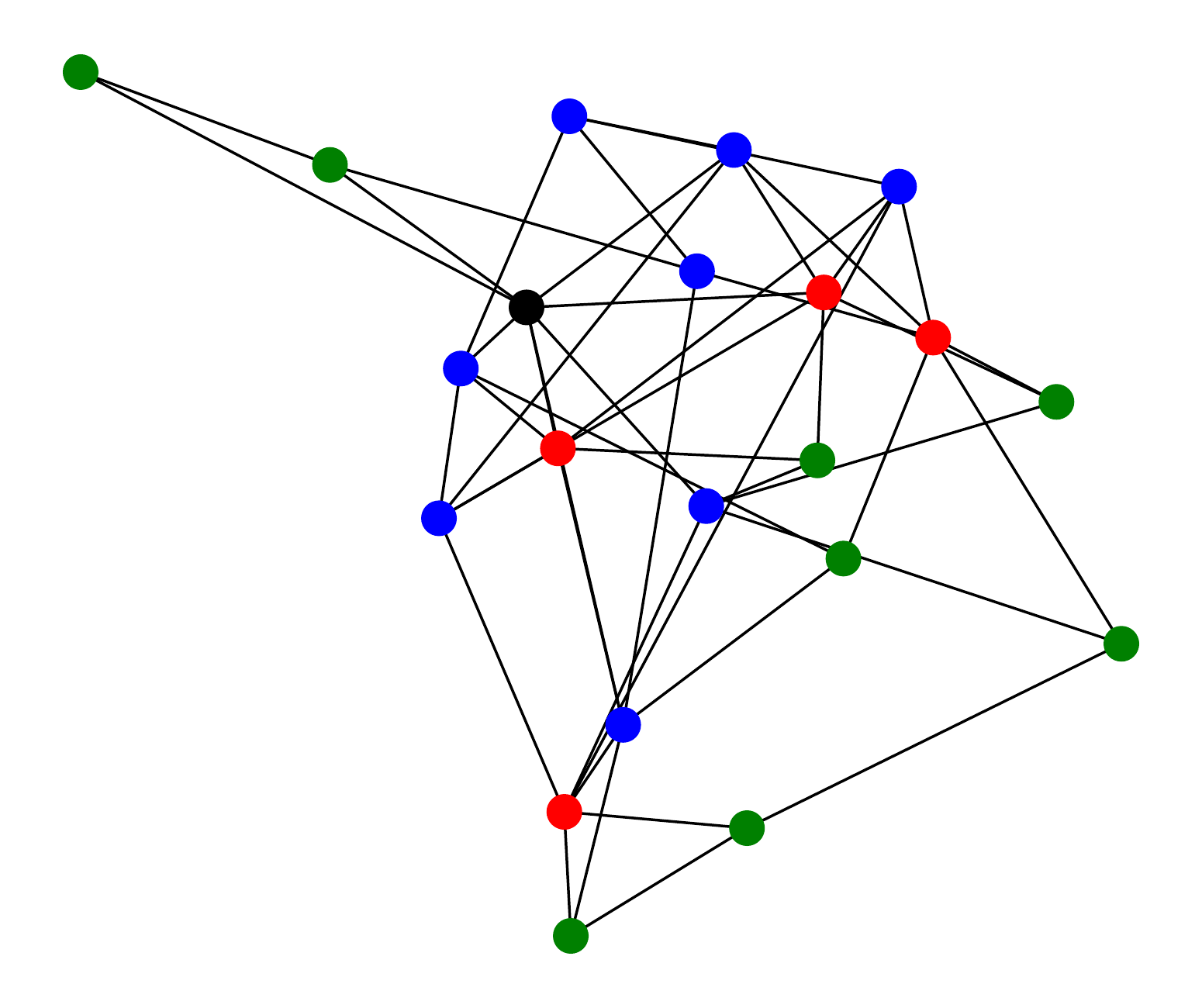}
		\caption{}
		\label{fig:topo_waxman}
	\end{subfigure}     
	\caption{\small{ \rev{The network graph representation for (a) N1 and (b) N2. The green, blue, red, and black dots represent the RUs, FSs, ESs and EPC, respectively. }   }}
	\vspace{-3mm}
	\label{fig:topo}
\end{figure}}

We consider a realistic MEC-based Milan topology (N1) \cite{milan_topology} and a synthetic topology (N2) generated using the Waxman algorithm \cite{waxman}, \rev{and their graph representation is illustrated in Fig. \ref{fig:topo}.}
N2 has parameters of link probability (0.5) and edge length control (0.1). A vRAN system in N1 and N2 consists of 1 EPC, 4 ESs, 8 FSs and 8 RUs (default), where the routers are co-located with each node\footnote{\rev{The datasets for N1 and N2 initially do not specify which nodes are for EPC, ESs, FSs, and RUs. We followed an intuitive approach by selecting them from the highest network degree.}}. Per link's latency, capacity, and weights of N1 and N2 vary from 0 to 0.1 ms, 30 Gbps to 160 Gbps, and 0 to 0.1.
We have $H_l = 20 \text{ RCs}, \forall l \in \mathcal{L}$ and $\hat{H}_m = 100 \text{ RCs}, \forall m \in \mathcal{M}$.
We set the available flavors with $|\mathcal{X}|=16$ for Platform A and Platform B, which translate to $\{0, 1, ...,14, 15\}$ RCs of the computing resources. 
Then, we define two vRAN systems in which we utilize Platform A with N1 (VR1) and Platform B with N2 (VR2). 

\rev{
We set the computing processing fee (per CPU usage) at the RU with $\kappa_\text{RU} \!\!:=\!\! 1 \text{ RC}^{-1}$ \cite{andres_fluidran_joint}. A single ES can serve up to 8 FSs, and a single FS can handle as high as 8 RUs. Therefore, we set $\kappa_\text{FS} :=  0.5 \kappa_\text{RU}$ and $ \kappa_\text{ES} := 0.5\kappa_\text{FS}$ (c.f. \cite[Fig. 6a]{complexity_cran} with $\approx 10$ BSs) by taking into account the processing gain from centralization (i.e., computational processing cost is less by centralizing more functions and executing them in a higher computing platform). Then, with regards to prior study in \cite{aztec}, we set the coefficient fee for resource overprovisioning with $\kappa_\text{O} := 1 \text{ RC}^{-1} $ and declined demands with $\kappa_\text{D} := 5 \text{ RC}^{-1}$. It is common that the penalty due to the declined demands incurs a higher cost. We also set the default coefficient for the reconfiguration fee lower with $\kappa_\text{R} = 0.1 \text{ RC}^{-1}$ to account for the typically relatively lower cost per unit of resource reconﬁguration \cite{aztec}. Then, we set $\kappa_\text{I} :=  \kappa_\text{R}$ (see Sec. \ref{sec:model}) and  $\kappa_\text{H} := 1 \text{ Gbps}^{-1}\text{/Km}$ (e.g., the fee for reserving 1 Gbps/Km routing bandwidth is the same as a processing fee at RU).} 

The Q-network of branching D3QN has an input layer with size of $|s|$, hidden layers (the architecture and size are provided in Fig. \ref{fig:bdq}), and $\sum_{k=1}^{K} C_k$ branches. Each branch has an output with size of $|\mathcal{A}_{kc}|$. The target network is updated every 500 time slots. The batch size is set with 128 and the replay buffer has a capacity of $10^6$. Our exploration and exploitation strategy is based on $\epsilon$-greedy, where we set $\epsilon_{\max}=1$ at the beginning of episode, then it exponentially decays to $\epsilon_{\min} = 0.015$. We use Adam optimizer \cite{adam_optim} with learning rate is set to $0.0001$ and \eqref{eq:loss_bdq} for the loss function. The time horizon for a single episodic training is one day ($N$ = 144 time slots) and the online operation starts on the second day with a default duration of two days ($N$ =  288 time slots). \rev{Table \ref{table:experimental_setup} summarizes the default experimental setups used in our evaluation.} The datasets in this work will be released online\footnote{\url{https://github.com/fahriwm/larv_datasets}}.

\begin{table}[t] \centering
	\begin{threeparttable}
		\begin{small}
			\rev{\begin{tabular}{@{}ll@{}}\toprule
					\textbf{Parameters}& \textbf{Default value}
					\\ \midrule
					Number of ESs ($M$) &      4
					\\ \hdashline
					Number of FSs ($L$) &      8
					\\ \hdashline
					Number of RUs ($K$) &      8
					\\ \hdashline
					FS computing capacity  (${H}_l$) &      20 RCs 
					\\ \hdashline
					ES computing capacity  ($\hat{H}_m$) & 100 RCs
					\\ \hdashline
					The set of flavors ($\mathcal{X}$) & $\{0,1,2, ..., 15\}$ RCs  
					\\ \hdashline
					Overprovisioning fee  ($\kappa_\text{O}$) &      $1 \text{ RC}^{-1}$
					\\ \hdashline
					Declined demand fee  ($\kappa_\text{D}$)  &      $5 \text{ RC}^{-1}$
					\\ \hdashline
					Reconfiguration fee  ($\kappa_\text{R}$) &      $0.1 \text{ RC}^{-1}$
					\\ \hdashline 
					Instantiation fee  ($\kappa_\text{I}$) &      $0.1 \text{ RC}^{-1}$
					\\ \hdashline 
					Processing fee at ES  ($\kappa_\text{ES}$) &      $0.25 \text{ RC}^{-1}$
					\\ \hdashline	
					Processing fee at FS  ($\kappa_\text{FS}$) &      $0.5 \text{ RC}^{-1}$
					\\ \hdashline 
					Processing fee at RU  ($\kappa_\text{RU}$) &      $1 \text{ RC}^{-1}$
					\\ \hdashline
					Bandwidth (routing) fee ($\kappa_\text{H}$) &      $1 \text{ Gbps}^{-1}\text{/Km}$
					\\ \hdashline
					Time horizon ($N$) for 1 episode  &      144 time slots 
					\\ \hdashline
					Epsilon start and end  ($\epsilon_{\max}, \epsilon_{\min}$)  &      (1, 0.015) 
					\\ \hdashline
					Learning rate  &      0.0001 
					\\ \hdashline
					Batch size  &      128
					\\ 
					\bottomrule
			\end{tabular}}
		\end{small}
	\end{threeparttable}
	\caption{\small\rev{  Experimental setup; see Sec. \ref{subsection:experimental_setup} for description.}}
	\label{table:experimental_setup}
	\vspace{-3mm}
\end{table}

\begin{figure*}[t!] \centering
	\begin{subfigure}[t]{.27\textwidth}
		\includegraphics[width=\textwidth]{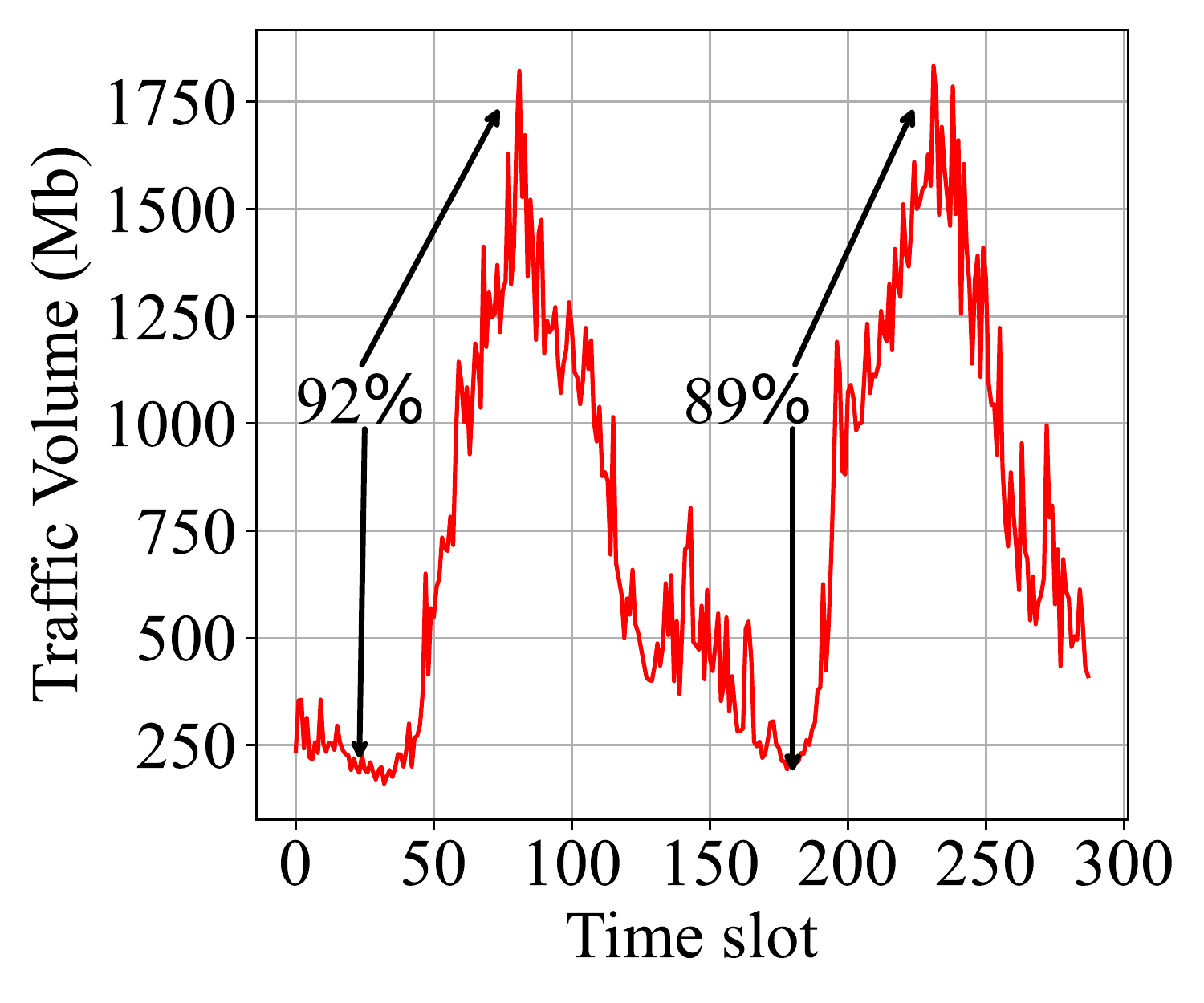}
		\caption{}
		\label{fig:dat_a}
	\end{subfigure} 
	\begin{subfigure}[t]{.27\textwidth}
		\includegraphics[width=\textwidth]{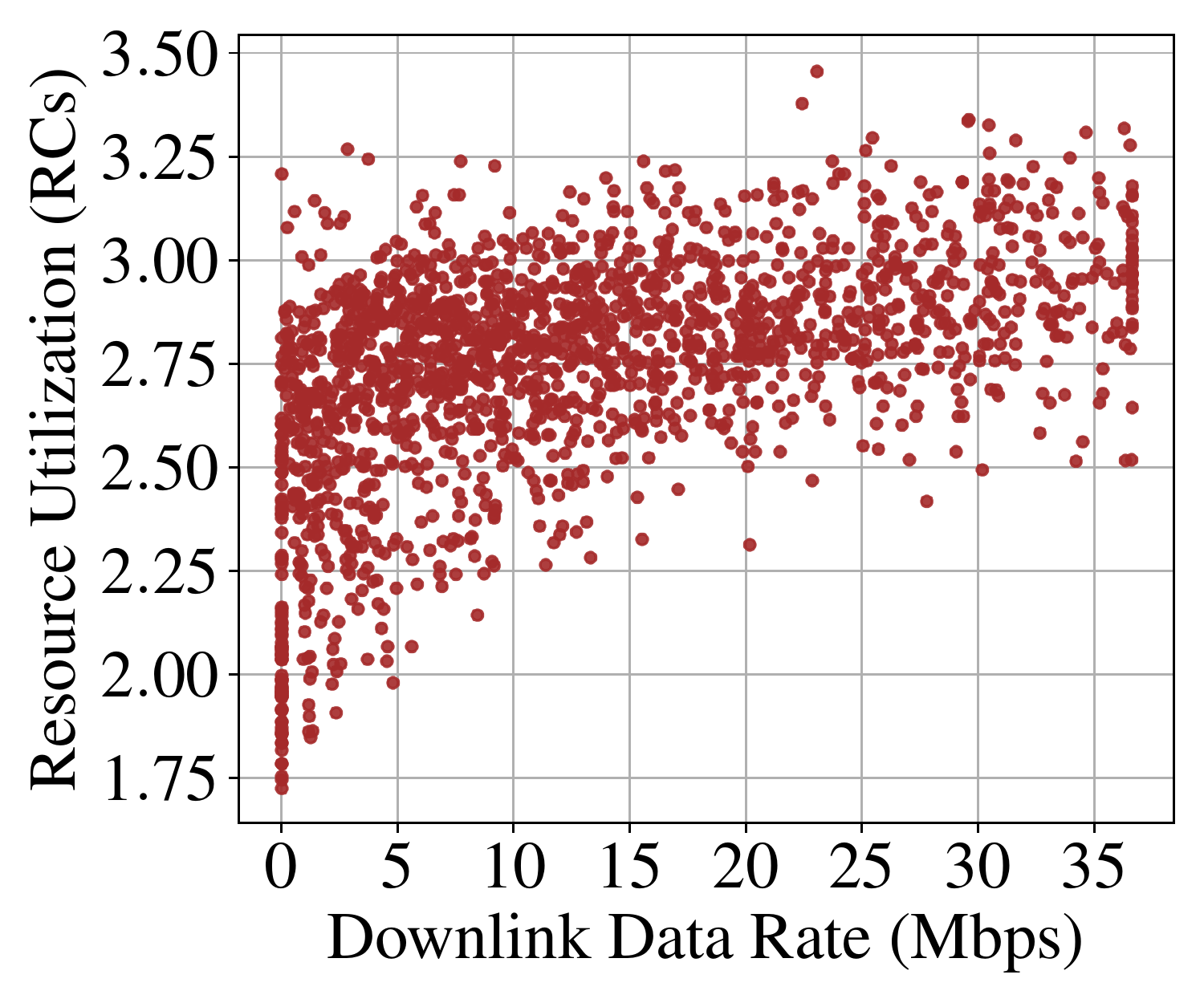}
		\caption{}
		\label{fig:dat_b}
	\end{subfigure}     
	\begin{subfigure}[t]{0.27\textwidth}
		\includegraphics[width=\textwidth]{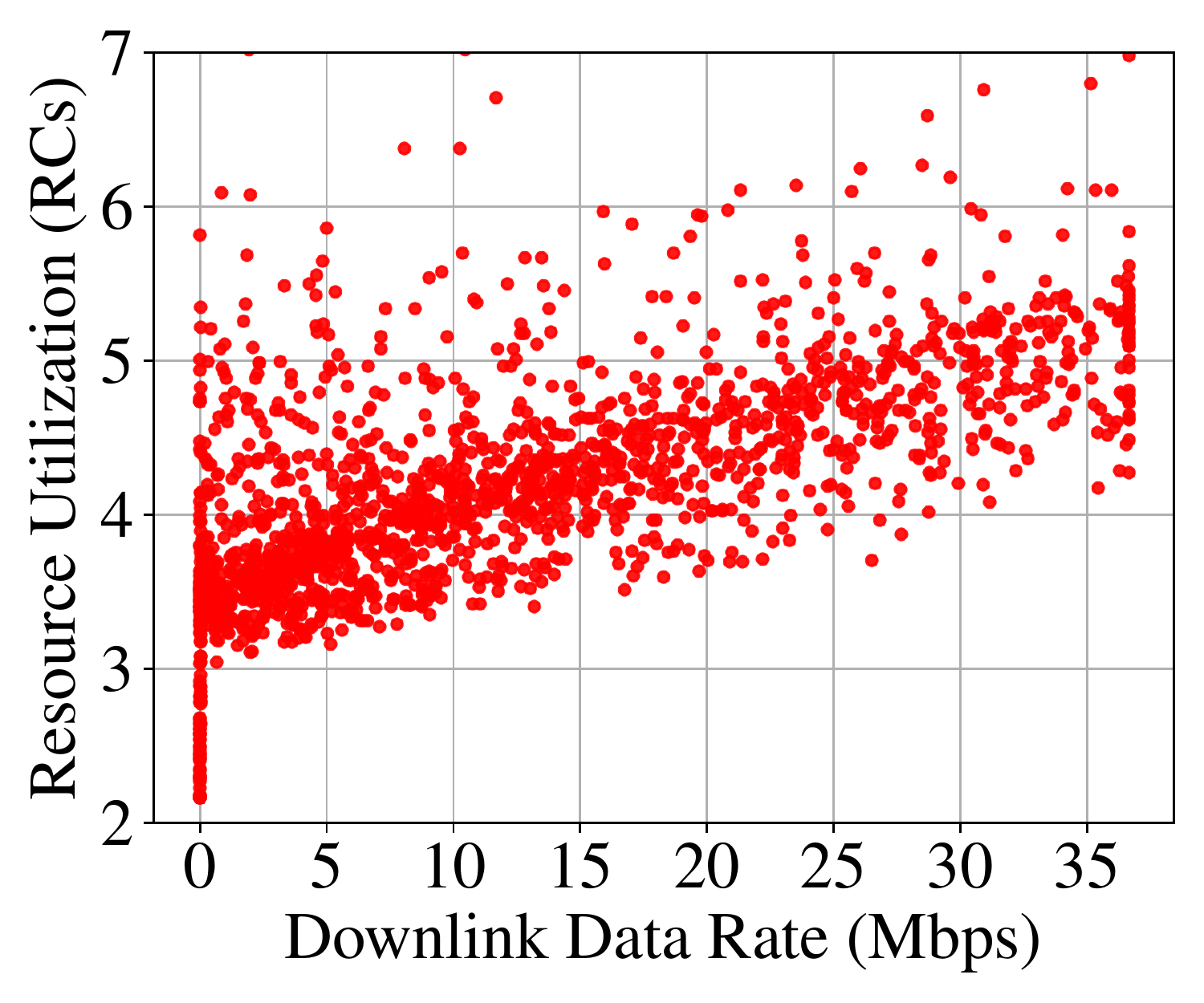}
		\caption{}
		\label{fig:dat_c}
	\end{subfigure} 
	\caption{\small{\textbf{a)} \rev{Traffic variation within two days from Milan datasets} \cite{milan_datasets} and \textbf{b)} collected measurement results over Platform A and \textbf{c)} Platform B. The resource utilization is presented in a reference core (RC), which translates to 1 virtual CPU/thread.
	}}
	\label{fig:data}
\end{figure*}

\begin{figure*}[t!]
	\centering
	\begin{subfigure}[t]{.4\textwidth}
		\includegraphics[width=\textwidth]{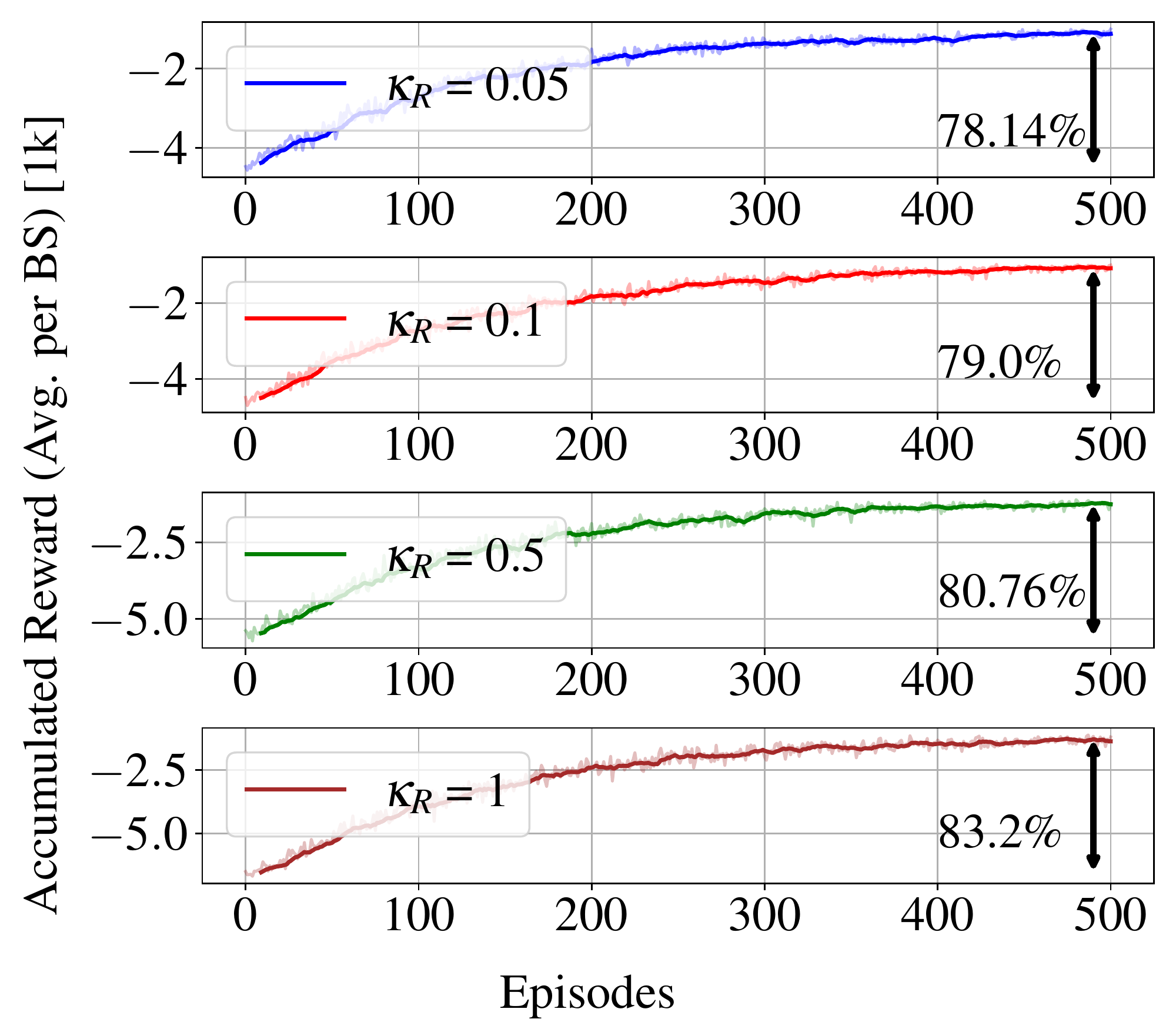}
		\caption{}
		\label{fig:train_reconfigure1}
	\end{subfigure} 
	\begin{subfigure}[t]{.4\textwidth}
		\includegraphics[width=\textwidth]{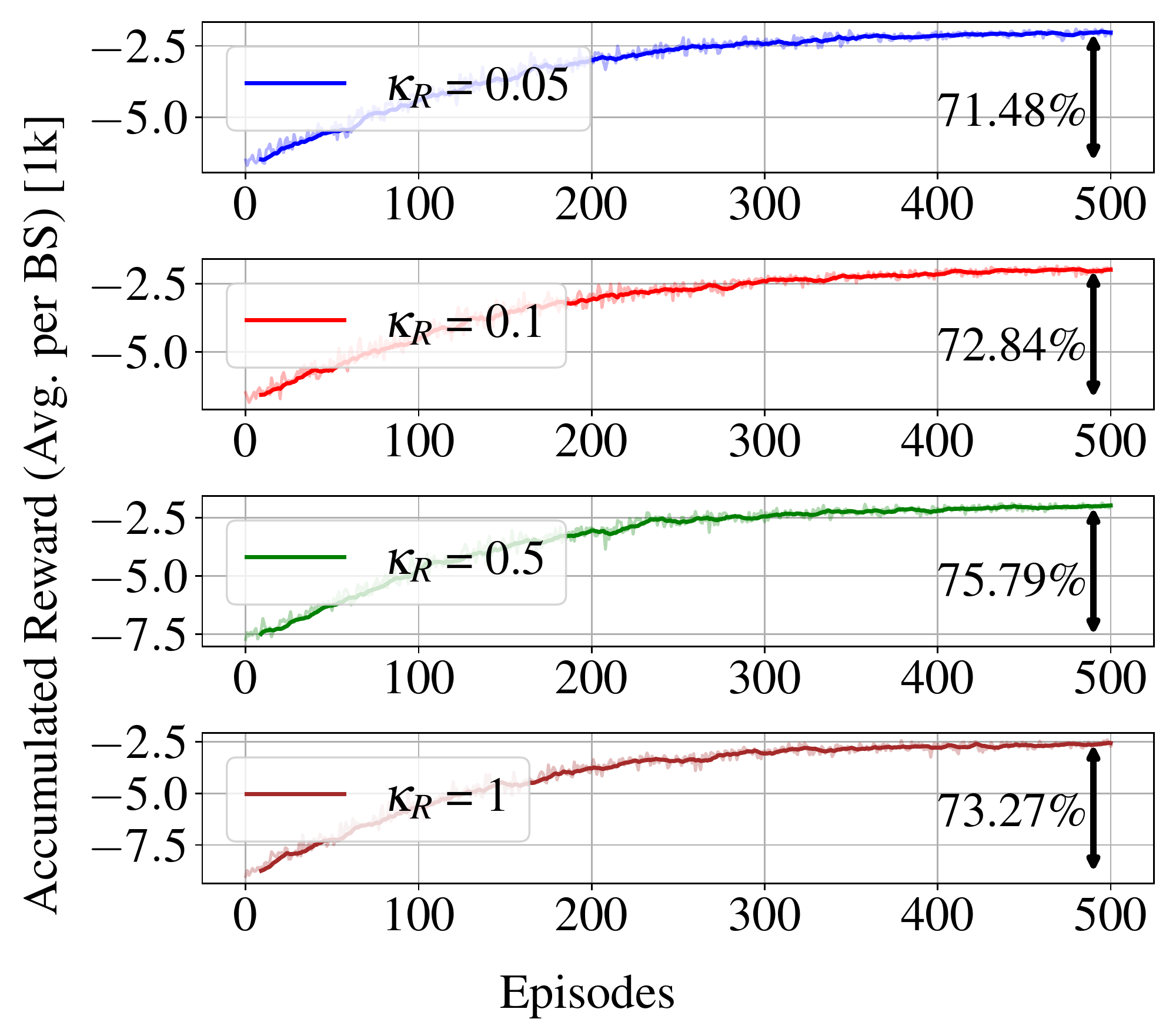}
		\caption{}
		\label{fig:train_reconfigure2}
	\end{subfigure}     
	\caption{\small{ The convergence of LARV under various reconfiguration fees in \textbf{a)} VR1 and \textbf{b)} VR2.   }}
	\vspace{-2mm}
	\label{fig:train_reconfigure}
\end{figure*}

Further, we compare LARV with several benchmarks as follows.
\begin{itemize}
	\item \textit{The best static with 100\% provisioning (BSP):} It knows exactly the peak future traffic demand of each BS and utilizes them to find the best static joint action via an exhaustive search. It can be defined as: $\pi_{BSP} := \arg\min_{a} \sum_{k}^K J_k^i (a)$, where $i = \arg\max_n \lambda_k^n$.
	Further, it is used to normalized the monetary costs in the online operation evaluations.
	
	\item \textit{DDPG with discretization:} Since the state space and action space of the RL problem are extremely large, the traditional discrete RL algorithm may not perform efficiently. A continuous RL algorithms such as DDPG \cite{ddpg} can address extremely large state-action space, but they are not designed for a discrete action. Hence, we relax the discrete action \eqref{eq:action_rl} into a continuous action. Then, when the output of DDPG is determined, we estimate it to the nearest discrete value. We also modify the output activation function with a Sigmoid function as each action needs to be a positive value.

	\item \textit{Multi-agent of D3QN (MDQ):} It is a non-branching D3QN. To deal with multi-dimensional action space, in every BS, each reconfiguration control works as a separate agent, i.e., the decision of each split, resource, and location is controlled by a different agent, that works collaboratively to maximize the common reward in \eqref{eq:reward_function_rl}. In total, MDQ has  $\sum_{k=1}^{K} C_k$ agents. The agents that represent control variables in the same BS share a common state observation. 
\end{itemize}

\vspace{-2mm}
\subsection{\rev{Measurement Insight}}

Fig. \ref{fig:dat_a} illustrates an example of the traffic demand of a BS in the Milan datasets \cite{milan_datasets}. It shows a significant difference between the peak and lowest traffic demand by up to 92\% in a single day. Moreover, the traffic variation might vary from day to day (e.g., weekdays, weekends). Figs. \ref{fig:dat_b} and \ref{fig:dat_c} show that the traffic demand highly affects the resource utilization of the BBU.
These findings motivate us to implement the dynamic configurations to adapt such traffic and resource variations to achieve cost-effective operations.
Figs. \ref{fig:dat_b} and \ref{fig:dat_c} also demonstrate that the relations between traffic demand and resource utilization have high variance, where we found a significant degree of spread on the resource utilization. Moreover, these relations are platform-dependent performance (e.g., hanging on the hosting platforms and platform load). For example, although they indicate not strongly linear in Platform A and B, the resulting Pearson coefficient is different with 0.513 and 0.654, respectively.
Also, albeit the BBU has been reserved with the same resources, Fig. \ref{fig:dat_c} shows that the BBU utilization of Platform B is higher than Platform A. Such platform-dependent performance is also found in \cite{vranai_journal} for uplink, where the computing behavior of vRANs is identified depending on many latent factors.

\vspace{-2mm}
\subsection{Performance during Training Process}
\begin{figure}[t]
	\centering
	\includegraphics[width=.4\textwidth]{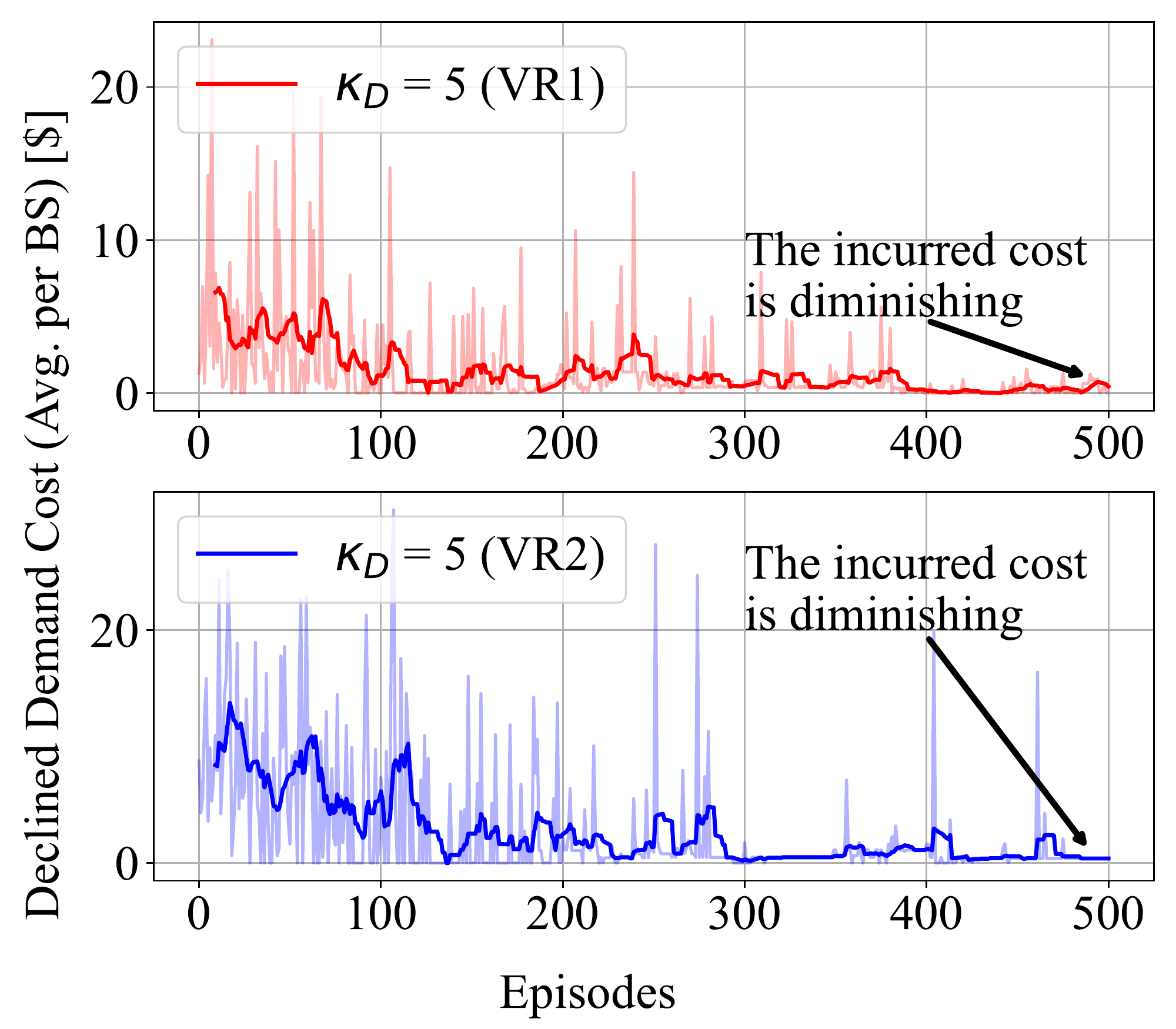}
	\caption{\small The incurred cost from declined service demands (average per BS) in VR1 and VR2. The cost is diminishing as the training goes, and it eventually reaches to near zero (e.g., after \rev{400} episodes).}
	\label{fig:train_declined}
	\vspace{-2mm}
\end{figure} 

\subsubsection{Training Convergence} Fig. \ref{fig:train_reconfigure} illustrates the convergence behavior of LARV over various reconfiguration coefficient fees in VR1 and VR2. \rev{At the beginning of episodes, LARV has a higher probability of utilizing a random policy for exploration. As a result, LARV produces a high long-term total operation cost over all the reconfiguration fees in VR1 and VR2. However, after some episodes, LARV successfully learns the optimal policy, starts to act greedily with a high probability, and convergences to the best policy the agent can learn.} Moreover, we found a similar trend in LARV's behavior, where it manages to converge to some cost values after 400 episodes, albeit it learns over different reconfiguration fees and vRAN systems.

Fig. \ref{fig:train_reconfigure} also shows that using a random policy in vRAN reconfiguration problem must be avoided as it yields in costly long-term cost.
In VR1, our findings reveal that LARV can save the costs by up to to 78.14\%, 79.0\%, 80.76\% and 83.2\% over  $\kappa_\text{R} = 0.05$,  $\kappa_\text{R} = 0.1$,  $\kappa_\text{R} = 0.5$ and  $\kappa_\text{R} = 1$, respectively, compared to a random policy. Such significant cost savings by LARV also appear in VR2, where LARV can save the cost as high as 75.79\%. The cost savings of LARV also increase when the reconfiguration fee is more expensive (e.g., $\kappa_\text{R} = 0.05$ to $\kappa_\text{R} = 1$).


\subsubsection{Declined Demands} Fig \ref{fig:train_declined} shows that LARV can reduce the incurred cost due to declined demands after several training episodes both in VR1 and VR2. The declined demand cost appears in almost every episode at the beginning of training episodes. \rev{The main reason is that LARV mostly chooses random actions for exploration, rendering a very high number of declined service demands and, at the same time, producing a very expensive cost. Note that the declined demands contribute a significantly more expensive cost as its coefficient fee is much higher than others.} 
As the training continues, LARV optimizes its weights based on the reward feedback and successfully diminishes the declined demand cost. After around 400 episodes, the incurred cost at each episode becomes smaller and less frequent, eventually reaching almost zero (or zero). Following the decrease of this cost, at the same time, the accumulated total operation cost (see Fig. \ref{fig:train_reconfigure}) is also greatly diminished.


%

\subsubsection{Transfer Learning} To assess the generalization of LARV over heterogeneous vRAN systems, we study the benefits of utilizing a transfer learning paradigm ("w/ transfer") compared to learning from scratch ("w/o transfer"). In particular, we leverage our pre-trained neural network weights (trained in VR1) for initializing the other neural network weights in different vRAN systems (e.g., in VR2). 
It is worth noting that the system parameters and platforms in VR1 and VR2 are different.
Hence, this evaluation \rev{aims} to study the possibility of reusing the existing models for the other vRAN systems, which might expedite the convergence and widespread deployment of LARV. 
We use the same default hyperparameter (defined in Sec. \ref{subsection:experimental_setup}), except we encourage less exploration for "w/ transfer" by modifying $\epsilon_{\text{max}} = 1$ to $\epsilon_{\text{max}} = 0.1$.

\begin{figure}[t!]
	\centering
	\includegraphics[width=.4\textwidth]{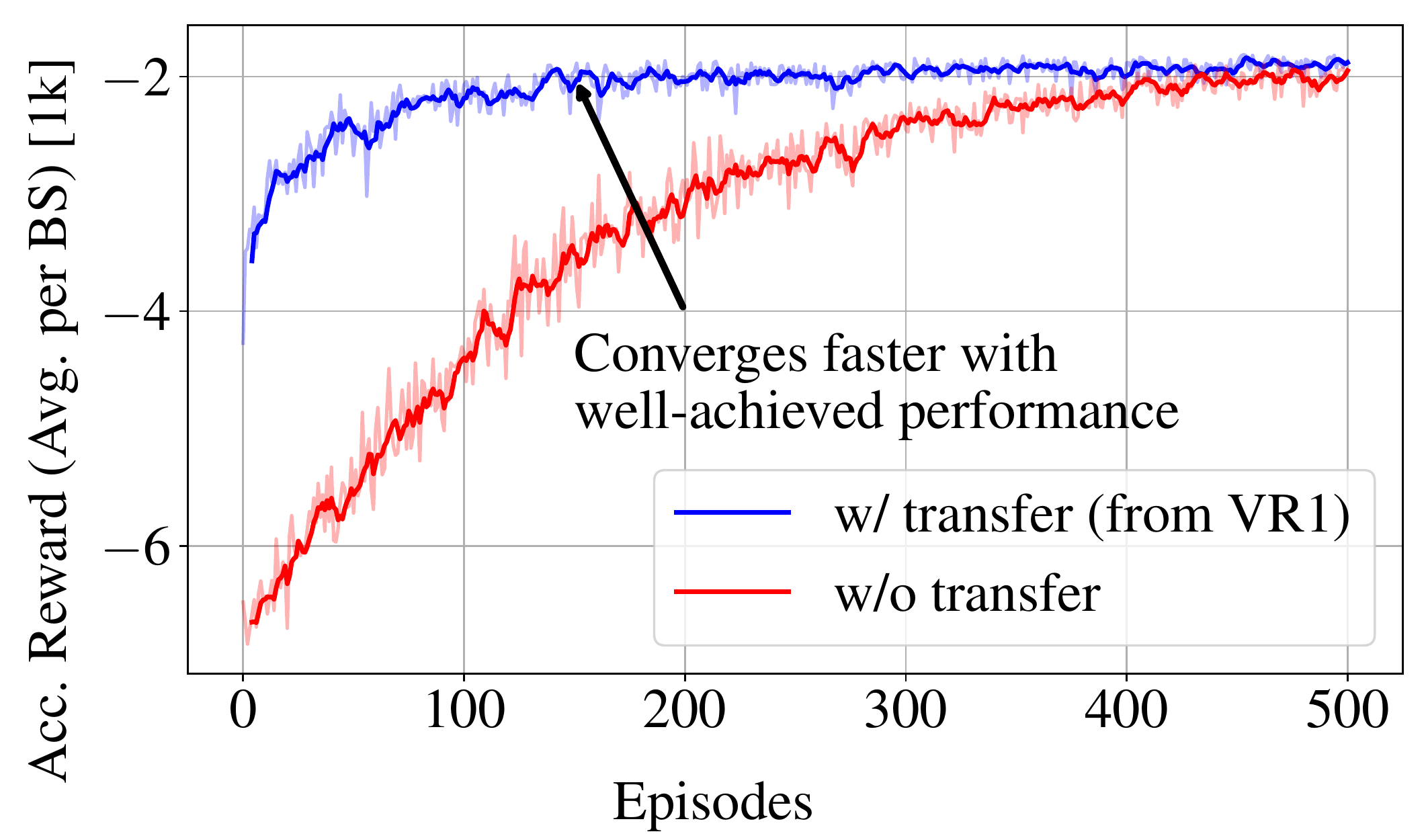}
	\caption{\small Training convergence in VR2. Using transfer learning paradigm ("w/ transfer"), which is leveraged from pretraining weights in VR1, can achieve similar performance and faster convergence compared to without transfer learning ("w/o transfer"). 
	}
	\label{fig:train_transfer}
	\vspace{-2mm}
\end{figure}   

Fig. \ref{fig:train_transfer} depicts that LARV "w/ transfer" successfully converges to the similar value with "w/o transfer" in VR2, albeit the pre-trained weights are leveraged from a different vRAN system (VR1).
Moreover, "w/ transfer" can speed up the training convergence with similar performance as "w/o transfer" even though the pre-training is conducted not in the same platform, where it starts to converge after around 150 episodes. 
\rev{In transfer learning, a pre-trained model is utilized. And when a pre-trained model is available, the gained knowledge of this already trained model can be transferred among \emph{different but similar (e.g., correlated)} environments and contexts, which in our case are VR1 and VR2. Such a transfer knowledge paradigm can expedite the learning convergence and allow the reuse of existing pre-trained models across different but related vRAN systems (i.e., have correlations with the training environment/context). }


\subsubsection{Action Space Compression}  \rev{Following the simulation setup, each BS has sub-action sizes with $|\mathcal{I}| \!=\! 4$, $|\mathcal{X}| \!=\! 16 $, $|\mathcal{L}| \!=\! 8$ and $|{\mathcal{M}}| \!=\! 4 $, and we have $|\mathcal{K}| = 8$. Hence, the number of Q values to be estimated is originally around $1.32 \times 10^{36}$. LARV turns such a combinatorial explosion into a linear increase; hence, the number of estimated Q values becomes 384.}

\begin{figure*}[t!]
	\centering
	\begin{subfigure}[t]{.4\textwidth}
		\includegraphics[width=\textwidth]{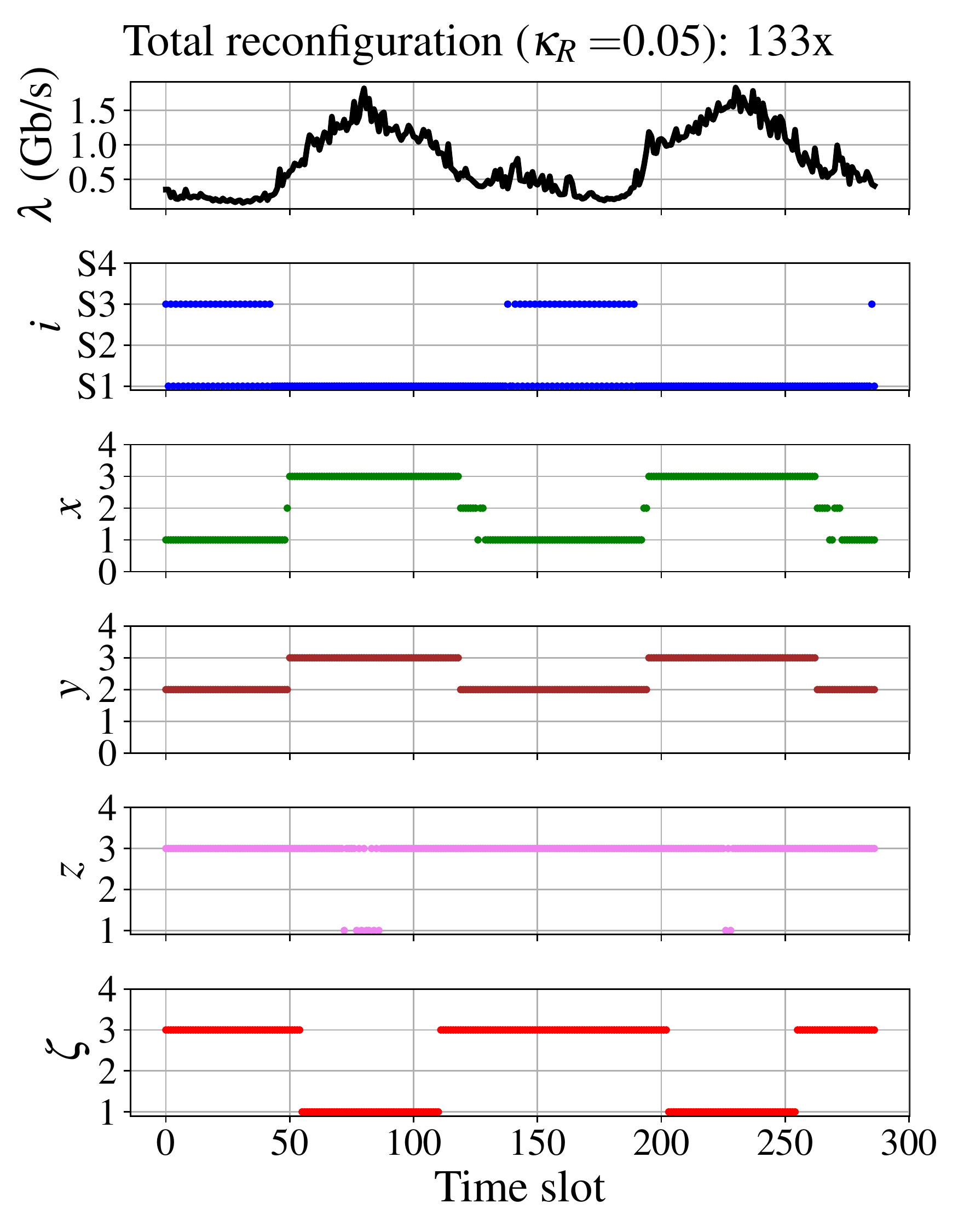}
		\caption{$\kappa_\text{R} = 0.05$}
		\label{fig:actions1}
	\end{subfigure} 
	\begin{subfigure}[t]{.4\textwidth}
		\includegraphics[width=\textwidth]{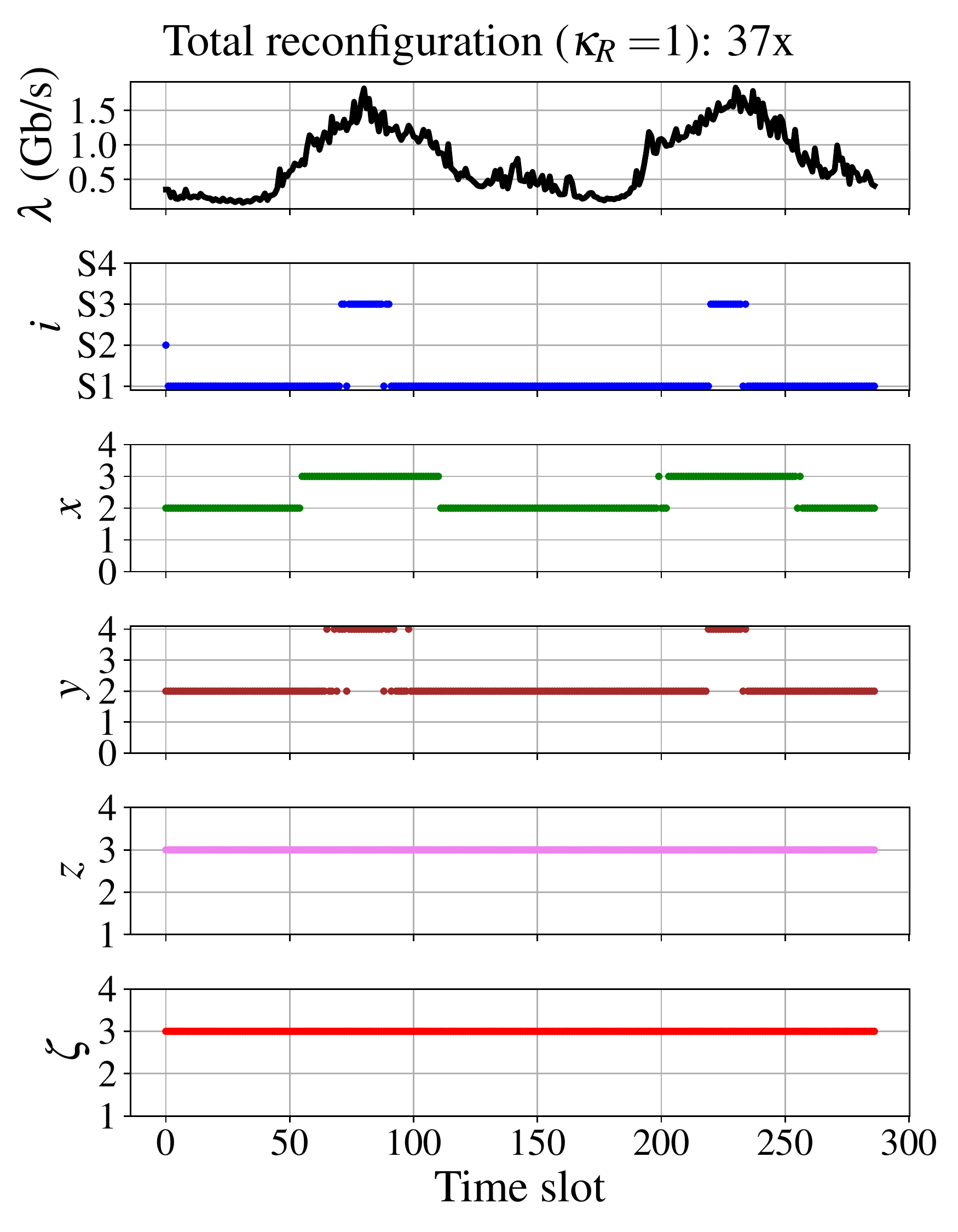}
		\caption{$\kappa_\text{R} = 1$}
		\label{fig:actions4}
	\end{subfigure}     
	\caption{\small{The selected actions over different reconfiguration fees and traffic demand variations at BS-1 during online operation (2 days).  }}
	\vspace{-2mm}
	\label{fig:actions}
\end{figure*}

\subsection{Performance during Online Operation}

\subsubsection{Selected actions}

Fig. \ref{fig:actions} illustrates how LARV successfullys controls the configurations of BS-1 reacting to the traffic variations and resource availability over different reconfiguration fees. \rev{Instead of minimizing the incurred cost at each slot, LARV's objective is to minimize the cost in the long run. As shown in Fig. \ref{fig:actions4}, LARV performs 133$\times$ reconfiguration activities when $\kappa_\text{R}=0.05$. However, this activity becomes less frequent with the increase of reconfiguration fee, where there are only 33$\times$ reconfiguration activities.} 
For the functional split ($i$), LARV mostly selects S1 (more decentralized functions) when the traffic of BS-1 is low, and it adjusts the split decision to S3 (more centralized functions) as the traffic increases. In S3, the transferred data flow over HLS is equal to the traffic demand with 500 Mbps of additional signaling overhead ($\lambda + 0.5$ Gbps); hence, LARV does not suggest implementing it in low traffic for such high overhead. However, when the traffic is elevated, i.e., the signaling does not significantly contribute to the data flow and routing cost, LARV tends to choose S3, considering the benefits of function centralization. 
This behavior appears in both  $\kappa_\text{R}=0.05$ and  $\kappa_\text{R}=1$, though the number of reconfigurations differs, where the reconfiguration is more often for $\kappa_\text{R}=0.05$. Further, the other results suggest that the allocated resources and the placement locations for vDUs and vCUs vary for different reconfiguration fees. For instance, the allocated resource of the vDU ($x$) in $\kappa_\text{R}=1$ is larger than in $\kappa_\text{R}=0.05$, even during the traffic is low, as LARV needs to accommodate the less frequent reconfigurations and more decentralized functions (it mostly implements S1). 
LARV also directly allocates a higher resource of the vCU ($y$) to avoid numerous reconfigurations when $\kappa_\text{R}=1$. Moreover, LARV decides to rarely reconfigure the vDU ($z$) and vCU ($\zeta$) locations or even does not reconfigure them when the fee is costly ($\kappa_\text{R}=1$), as altering such configurations requires migrating all the resources to the new places, which can trigger significantly expensive reconfiguration cost.


\begin{figure*}[t!]
	\centering
	\begin{subfigure}[t]{0.36\textwidth}
		\includegraphics[width=\textwidth]{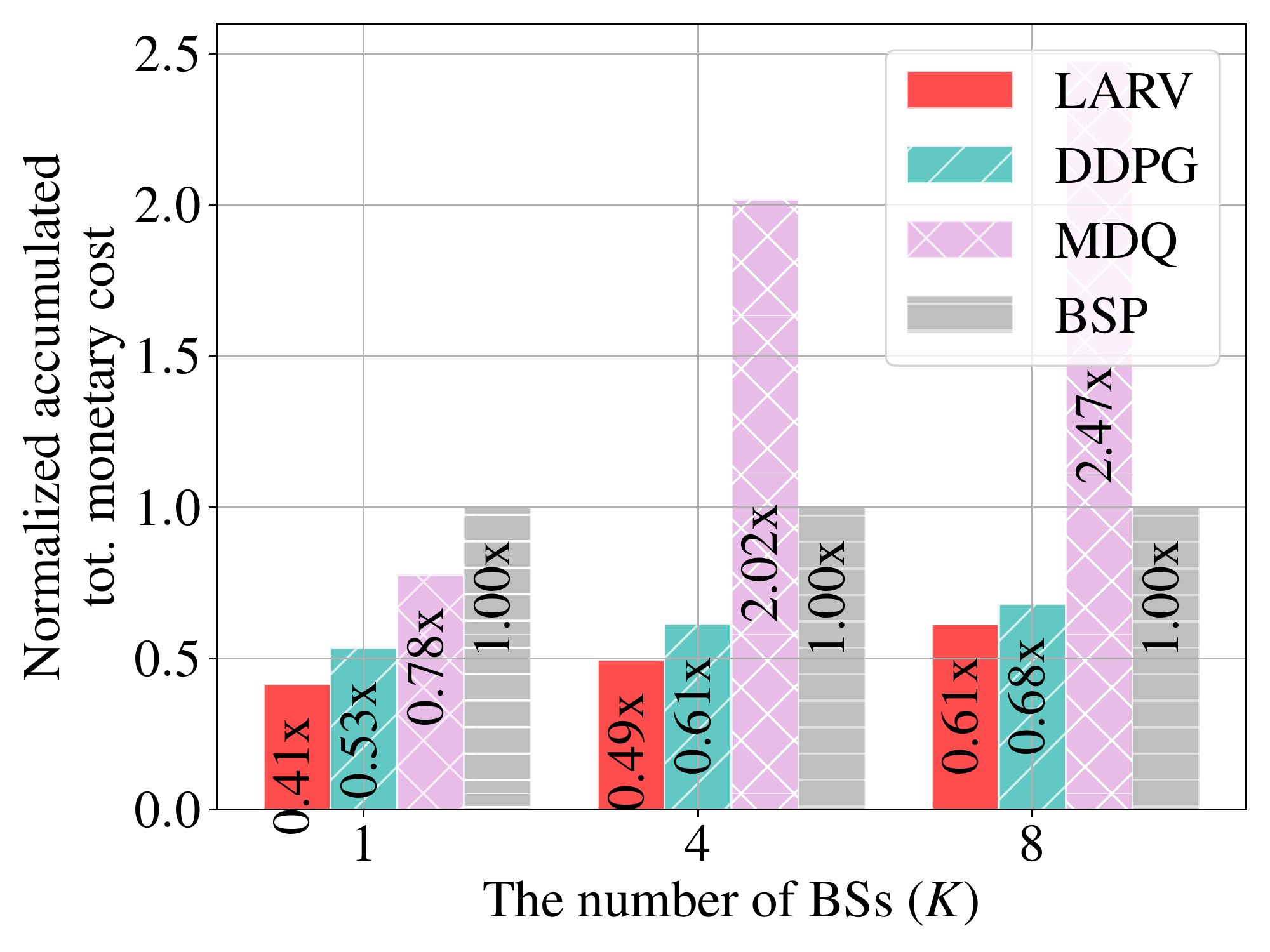}
		\caption{}
		\label{fig:online_nbs}
	\end{subfigure}
	\begin{subfigure}[t]{.36\textwidth}
		\includegraphics[width=\textwidth]{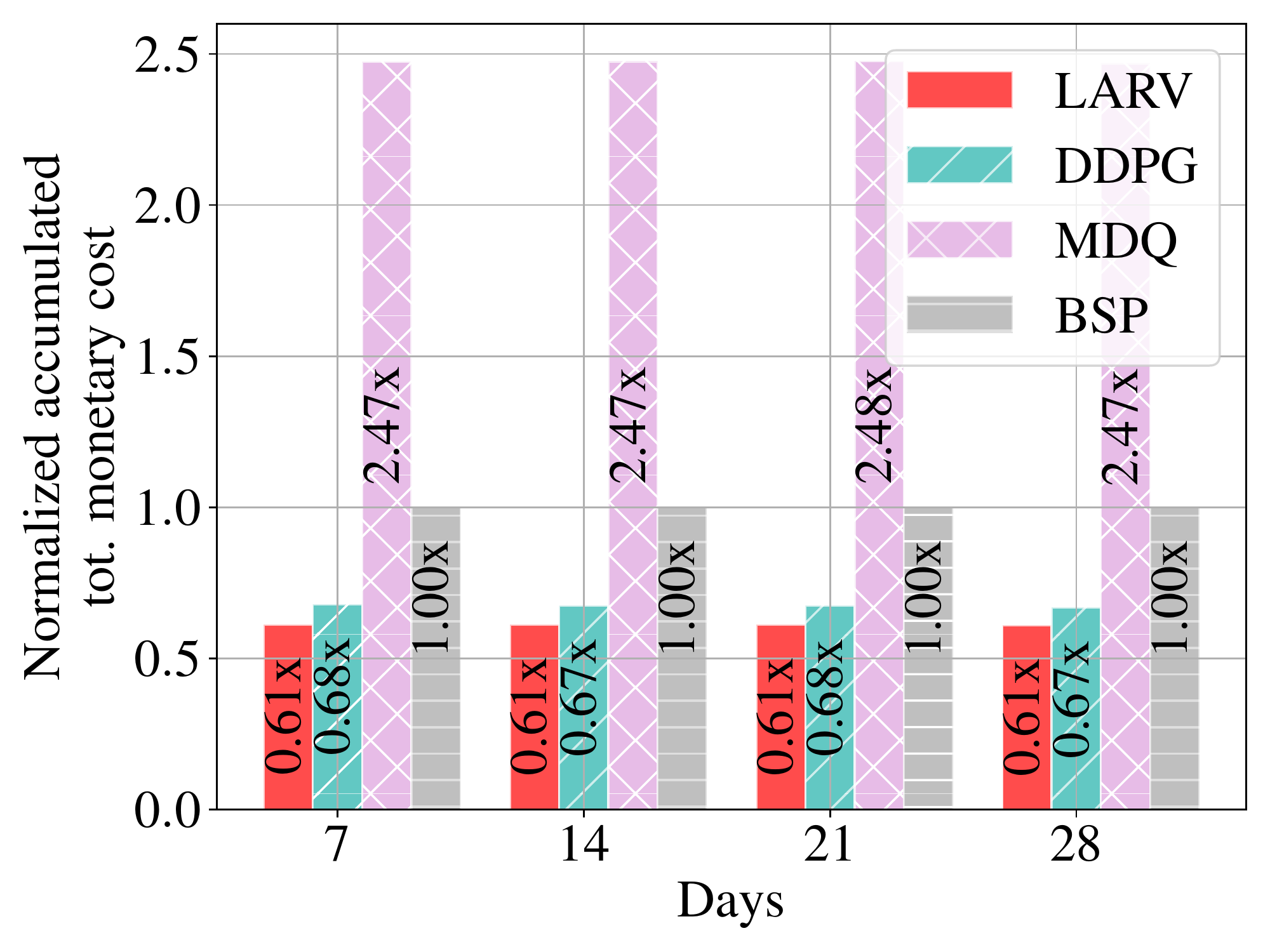}
		\caption{}
		\label{fig:online_timeh}
	\end{subfigure} 
	\begin{subfigure}[t]{.36\textwidth}
		\includegraphics[width=\textwidth]{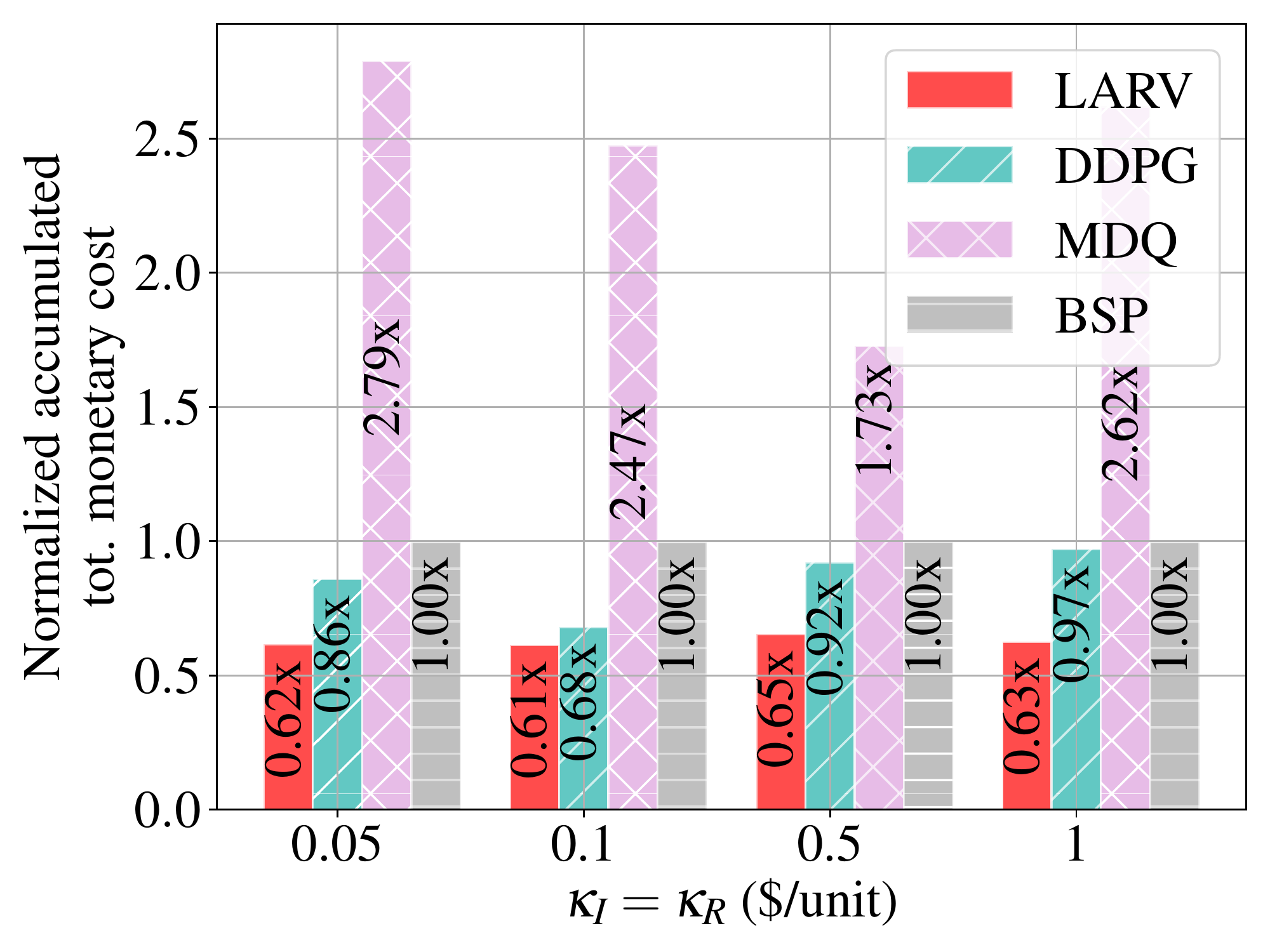}
		\caption{}
		\label{fig:online_reconfiguration}
	\end{subfigure}     
	\begin{subfigure}[t]{0.36\textwidth}
		\includegraphics[width=\textwidth]{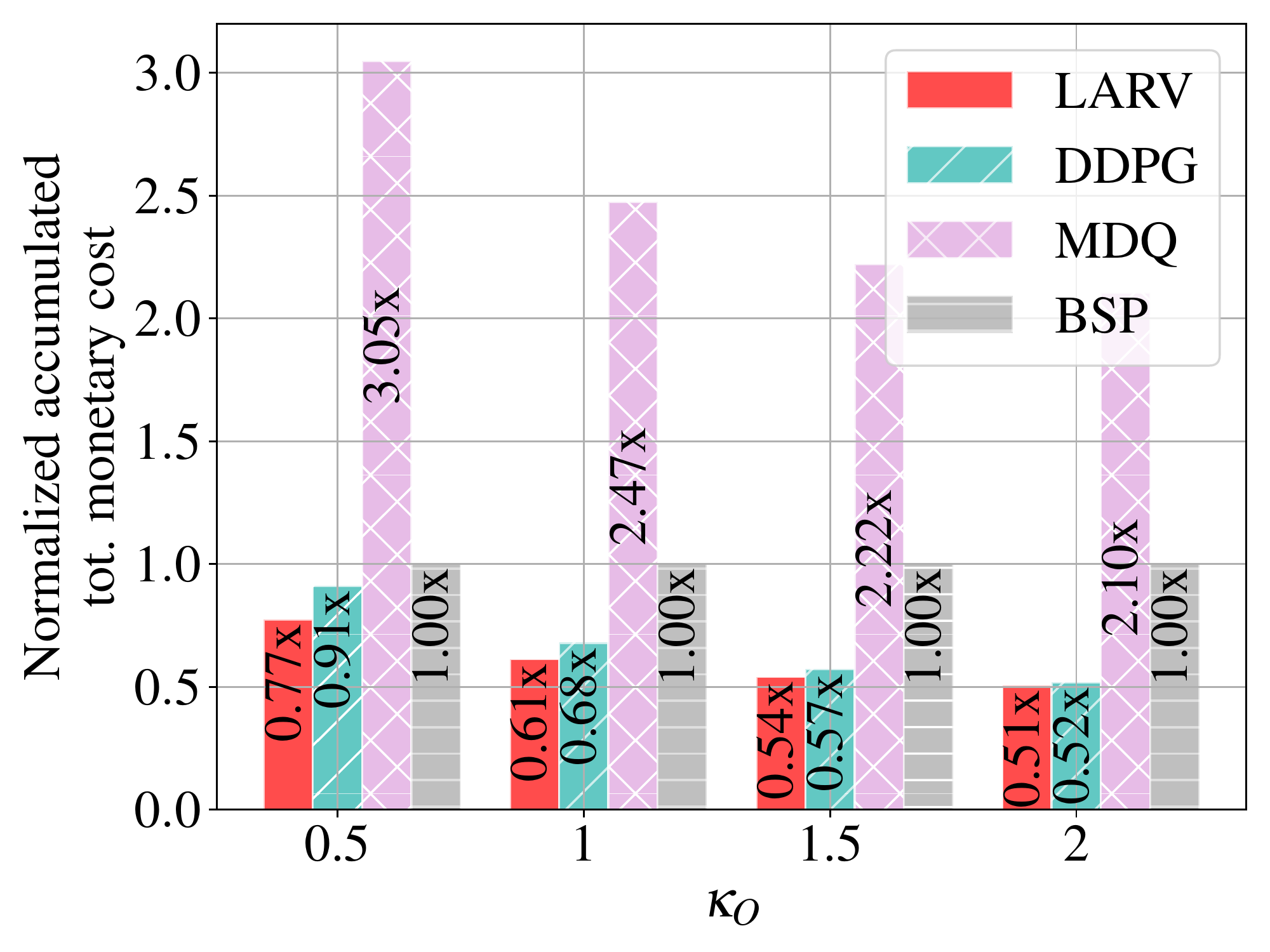}
		\caption{}
		\label{fig:online_overprovisioning}
	\end{subfigure}
	\caption{\small{Performance during the online operation in VR1. The presented monetary costs are normalized to BSP.  }}
	\vspace{-2mm}
	\label{fig:online}
\end{figure*}

\subsubsection{The Number of BSs} We evaluate LARV over a different number of the BSs in the vRAN system and present it in Fig. \ref{fig:online_nbs}. The number of BSs significantly influences the size of the state space and action space of the RL problem.
In general, all the RL approaches outperform BSP when $K=1$, where LARV becomes the most cost-effective by saving the cost up to 59\%. However, when the number of BSs in the vRAN system becomes more prominent, the size of the action space, state space, and the number of possible actions grow combinatorially. \rev{By adopting action branching, LARV successfully deals with such a combinatorial growth with a linear increase, rendering well-achieved performance, as shown in Fig. \ref{fig:online_nbs}. And it brings LARV to be the least degraded performance,} where the cost savings of LARV is more than 39\% of BSP. In contrast to LARV, MDQ utilizes a distributed multi-agent system. When the number of BSs increases, the number of agents of MDQ also increases, and this makes the performance of MDQ deteriorate compared to the centralized learning approaches. Moreover, albeit DDPG can deal with discrete action space through discretization of continuous action, the performance is still far from LARV. Unlike LARV, which is naturally designed for large discrete action space, DDPG can lose its learning effectiveness due to discretization.


\subsubsection{\rev{Time horizon setting}} Fig. \ref{fig:online_timeh} visualizes the performance of LARV compared to the benchmarks over various time horizon settings, ranging from 7 days to 28 days. We found that LARV becomes the most cost-effective approach by having the cheapest long-term total cost. The performance of LARV also remains stable, albeit in varying conditions (demands and resource availability). Compared to BSP, the cost-savings of LARV can be as high as 39\%. 
\rev{LARV updates the vRAN configurations prudently, adapting to the varying conditions and considering the long-term cost, while BSP follows static policy by using future traffic information. This finding clearly emphasizes the importance of dynamic reconfiguration in vRANs.} Moreover, LARV also outperforms RL benchmarks, where it saves the long-term total cost by up to 10\% of DDPG and 75\% MDQ. \rev{Compared to continuous space and non-branching state-of-the-art deep RL approaches, this gain shows the effectiveness of LARV through branching of D3QN in solving a large state space and multi-dimensional action space of the vRAN reconfiguration problem.}


\subsubsection{Reconfiguration fees} We analyze the impact of various reconfiguration fees $(\kappa_\text{R})$ on the cost savings that LARV can achieve. Fig. \ref{fig:online_reconfiguration} shows that LARV can successfully provide well-achieve performance in both cheap and expensive reconfiguration fees. It also shows that the increase in reconfiguration fee slightly affects the performance of LARV while it significantly degrades DDPG. \rev{DDPG is proposed for continuous action, and the performance can be deteriorated due to discretization when the problem has discrete action space, such as arising in our problem.} In general, compared to DDPG, the cost savings of LARV increase as the fee gets more expensive, where the gains of LARV rise from 10\% to as high as 35\% ($\kappa_\text{R} = 1$). 
Moreover, the cost savings of LARV remain stable compared to BSP and MDQ at around 35-39\% and 62-76\%, respectively. These findings emphasize that reconfiguring the vRAN system is beneficial, but we need to carefully design the RL algorithm suited to the vRAN problem.


\subsubsection{Overprovisioning fees} We study the effect of different overprovisioning coefficient fees to the performance of LARV. 
As seen from Fig. \ref{fig:online_overprovisioning}, when the overprovisioning fee gets costly, all the RL approaches' performance increases correspond to the static policy, where LARV becomes the best approach among them. LARV can save the costs from 23\% ($\kappa_\text{O}= 0.05$) to 49\% compared to BSP ($\kappa_\text{O}= 1$). 
These results highlight the need for reconfiguring \rev{prudently} the vRAN system at runtime, particularly when the resources are valuable and the price of wasting such resources is high, making the static policy economically unviable for long-term operations.


%
%

%% file: Conclusions.tex
\section{Conculusion} \label{sec:conclusion}

In this paper, we have proposed LARV that jointly reconfigures the functional splits of the BSs, the resources and placements of vDUs and vCUs, and the routing for each BS flow. The objective of LARV is to minimize the long-term total operation cost while adapting to the possibly-varying traffic demands and resource availability. In particular, we have analyzed the relations between the traffic demands and resource utilization in the vRAN system, which renders their relations have high variance and dependence on platform and platform load. We also have formulated a comprehensive cost model capturing the impacts of resource overprovisioning, instantiation and reconfiguration and the declined demands. We have developed LARV using a model-free deep RL paradigm to solve the sequential decision-making problem. The agent's neural network is developed using a combination of D3QN and action branching to tackle the large state space and multi-dimensional action space.
We also have conducted a series of trace-driven evaluations during the training process and online operation. The numerical results have shown that LARV successfully learns the optimal policy, \rev{where its learning convergence can be expedited through transfer learning even in different vRAN systems}. Moreover, LARV offers considerable cost savings by up to 59\% of the static benchmark,  35\% of DDPG with discretization, and 76\% of a distributed non-branching D3QN solution. 

\rev{ The proposed framework in this paper has been evaluated in a realistic simulated vRAN system based on collected testbed traces and network datasets. However, it has not been implemented in a real live network due to the limitation of the current testbed setup, i.e., it could not support several functional splits and the geographical location of the servers. In the future, implementing the framework and evaluating its performance in a real live network setup would be an interesting study.  }
%